\documentclass{article}
\usepackage{amsmath,amsfonts,amsthm}
\usepackage{amstext}
\usepackage{amsgen}
\usepackage{amsbsy}
\usepackage{amsopn}
\usepackage{amssymb}
\usepackage{graphicx}
\usepackage{epsfig}
\usepackage[ampersand]{easylist}
\newtheorem{thm}{Theorem}[section]

\newtheorem{algorithm}[thm]{Algorithm}
\newtheorem{proposition}[thm]{Proposition}
\theoremstyle{definition}

\theoremstyle{remark}

\numberwithin{equation}{section}\newcommand{\Real}{\mathbb{R}}

\newcommand{\msp}{\hspace{0.5cm}}

\def\IMAGESPATH{.//}

\begin{document}

\title{Discrete Optimal Global Convergence of an Evolutionary Algorithm
for Clusters under the  Potential of Lennard Jones}
\author{Carlos Barr\'{o}n-Romero \\
%EndAName
cbarron@correo.azc.uam.mx\\
\\
Universidad Aut\'onoma Metropolitana, Unidad Azcapotzalco \\
Av. San Pablo No. 180, Col. Reynosa Tamaulipas, C.P. 02200, \\
MEXICO }
\date{2016}
\maketitle

% ----------------------------------------------------------------
\begin{abstract}

A review of the properties that bond the particles under Lennard Jones Potential allow to states properties and conditions for building evolutive algorithms using the CB lattice with other different lattices. The new lattice is called CB lattice and it is based on small cubes, such the number of its vertices in a region is always greater than the number of the particles of a cluster or a region of a lattice inside of the same size region of the CB lattice. Moreover, the estimation of a putative optimal cluster of the Lennard Jones can be done theoretically in short time but, the proof, for such cluster to be the global optimal cannot be determining in efficient time. The proof of the global optimality for a cluster is related to the binomial coefficient $\binom{m}{n}$, which it corresponds with the selection of $n$ particles from a collection with $m$ given particles. A set of propositions states convergence and optimal conditions over the CB lattice for an evolutionary algorithm. The evolutionary algorithm is a reload version of previous genetic algorithms based in phenotypes. The novelty using CB lattice, together with the other lattices, and ad-hoc cluster segmentation and enumeration, is to allow the combination of genotype (DNA coding for cluster using their particle's number) and phenotype (geometrical shapes using particle's coordinates in 3D). A parallel version of an evolutionary algorithm for determining the global optimality is depicted. The algorithm for determining global optimality (which it is far from this research, and it is not included) is just a force brute searching algorithm with complexity $\binom{m}{n}$, where $n$ is the number of the cluster's particles and $m \gg n$ is the number of particles of an appropriate CB lattice's region. The results presented are from a standalone program for a personal computer of the evolutionary algorithm, which can estimate all putative Optimal Lennard Jones Clusters from 13 to 1612 particles. The novelty are the theoretical results for the evolutionary algorithm's efficiency, the strategies with phenotype or genotype, and the classification of the clusters based in an ad-hoc geometric algorithm for segmenting a cluster into its nucleus and layers. Also, the standalone program is not only capable to replicate the optimal Lennard Jones clusters in The Cambridge Cluster Database (CCD), but to find new ones.

\end{abstract}

%%%%%%%%%%%%%%%%%%%%%%%%%%%%%%%%%%%%%%%%%%%%%%%%%%%%%%%%%%%%%%%%%%%%%%%%%%%
%%%%%%%%%%%%%%%%%%%%%%%%%%%%%%%%%%%%%%%%%%%%%%%%%%%%%%%%%%%%%%%%%%%%%%%%%%%
%%%%%%%%%%%%%%%%%%%%%%%%%%%%%%%%%%%%%%%%%%%%%%%%%%%%%%%%%%%%%%%%%%%%%%%%%%%
\section{Introduction}
%%%%%%%%%%%%%%%%%%%%%%%%%%%%%%%%%%%%%%%%%%%%%%%%%%%%%%%%%%%%%%%%%%%%%%%%%%%

The problem for determining optimal clusters under Lennard Jones
captures my attention for the possible implications for building
an efficient algorithm for the class of NP.  My
techniques for  the NP Class has an application for building an
appropriate algorithms for looking the optimal
clusters under Lennard Jones Potential.

Over a decade ago, I states the conjecture
in~\cite{arXiv:Barron2005} that IF lattice could contain all optimal
clusters under the Potential of Lennard Jones (LJ). The title of
the article: Minimum search space and efficient methods for
structural cluster optimization was proposed as result of some
inquiries from D. J. Wales, J. P. K. Doye,  G.L. Xue and Bern
Hardke about the optimal LJ clusters.

\begin{figure}
\begin{center}
\psfig{figure=\IMAGESPATH/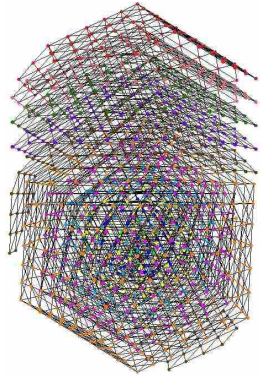,height=50mm}
\caption{MIF1739 contains the initial particles' positions for the $C^*_n,$
 $n=2,\ldots,
1000.$}~\label{fig:min_lattice1739}
\end{center}
\end{figure}

\begin{figure}
\centerline{ \psfig{figure=\IMAGESPATH/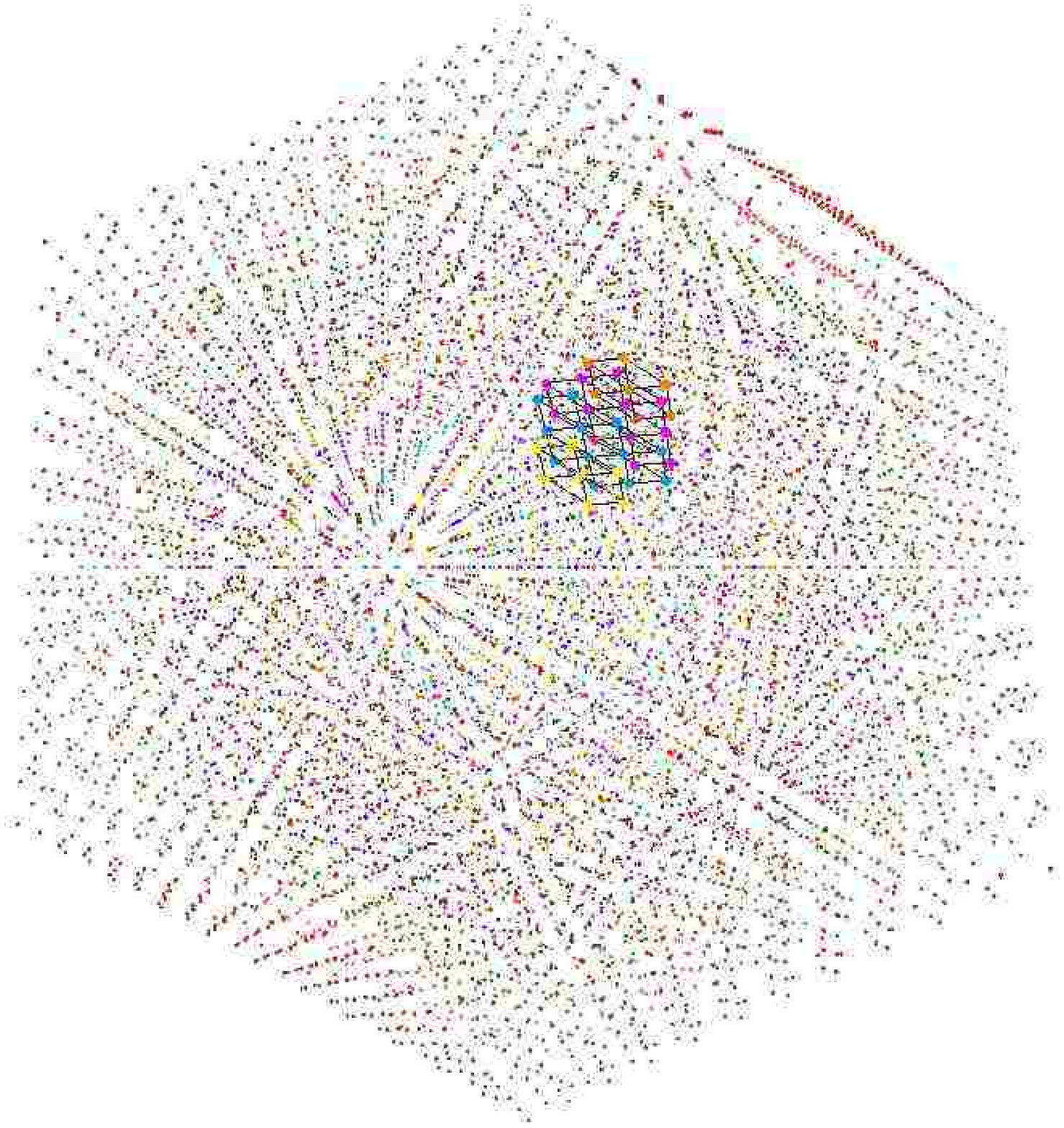,
height=50mm} \psfig{figure=\IMAGESPATH/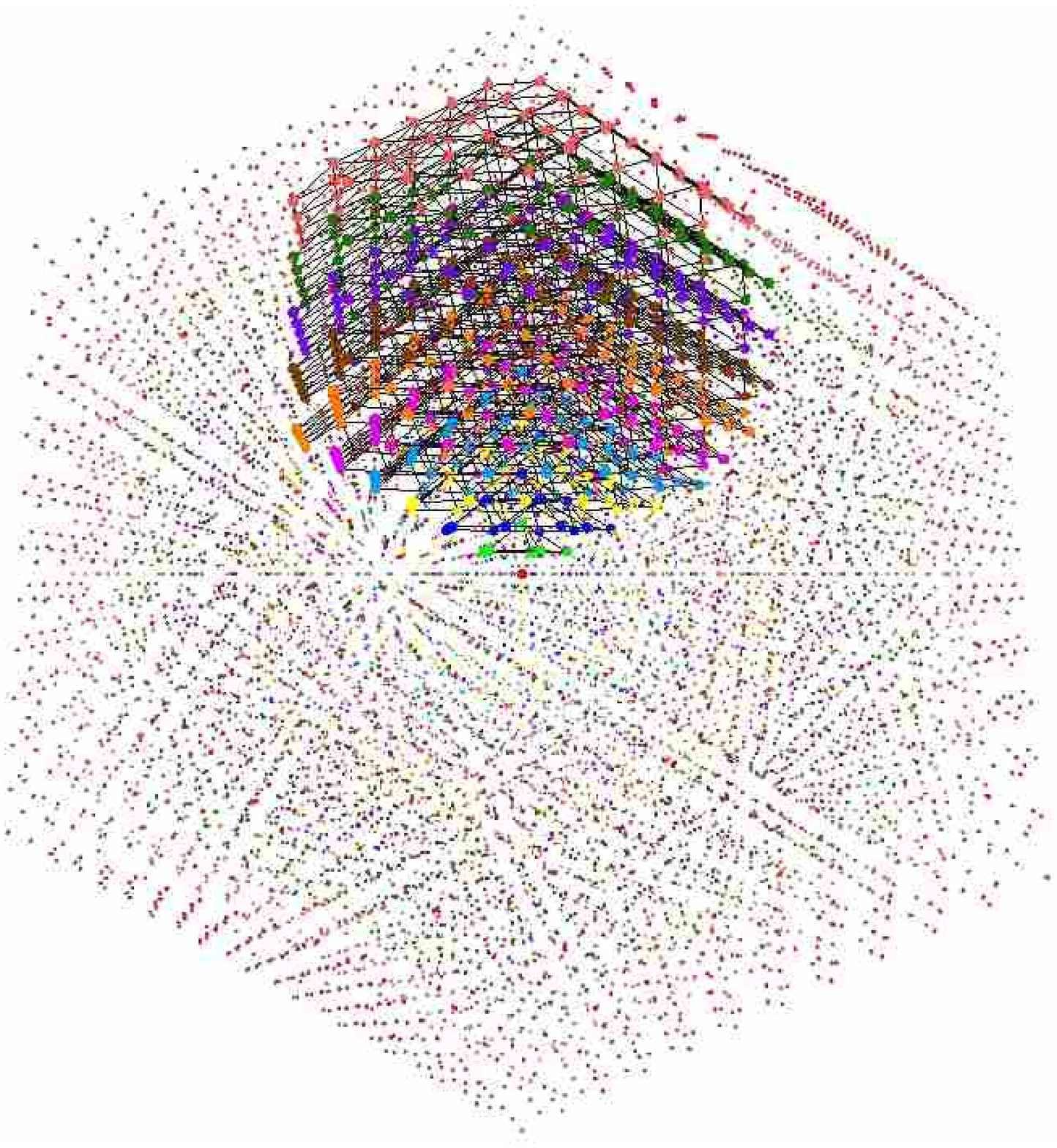,
height=50mm}} \centerline{ \makebox[2.2in][c]{ a)
}\makebox[2.2in][c]{ b) } } \caption{a) $C^*_{38}$ and b) $C^*_{664}$ inside of a region of the IF
lattice.}~\label{fig:LJ38:664}
\end{figure}

IF lattice results from overlapping the positions of the IC
lattice and FC lattice. The main result was a minimum region of
IF, where all putative global optimal LJ clusters from 2 to 1000
can be found (see figures~\ref{fig:min_lattice1739}, and
\ref{fig:LJ38:664}).

\begin{figure}
\centerline{ \psfig{figure=\IMAGESPATH/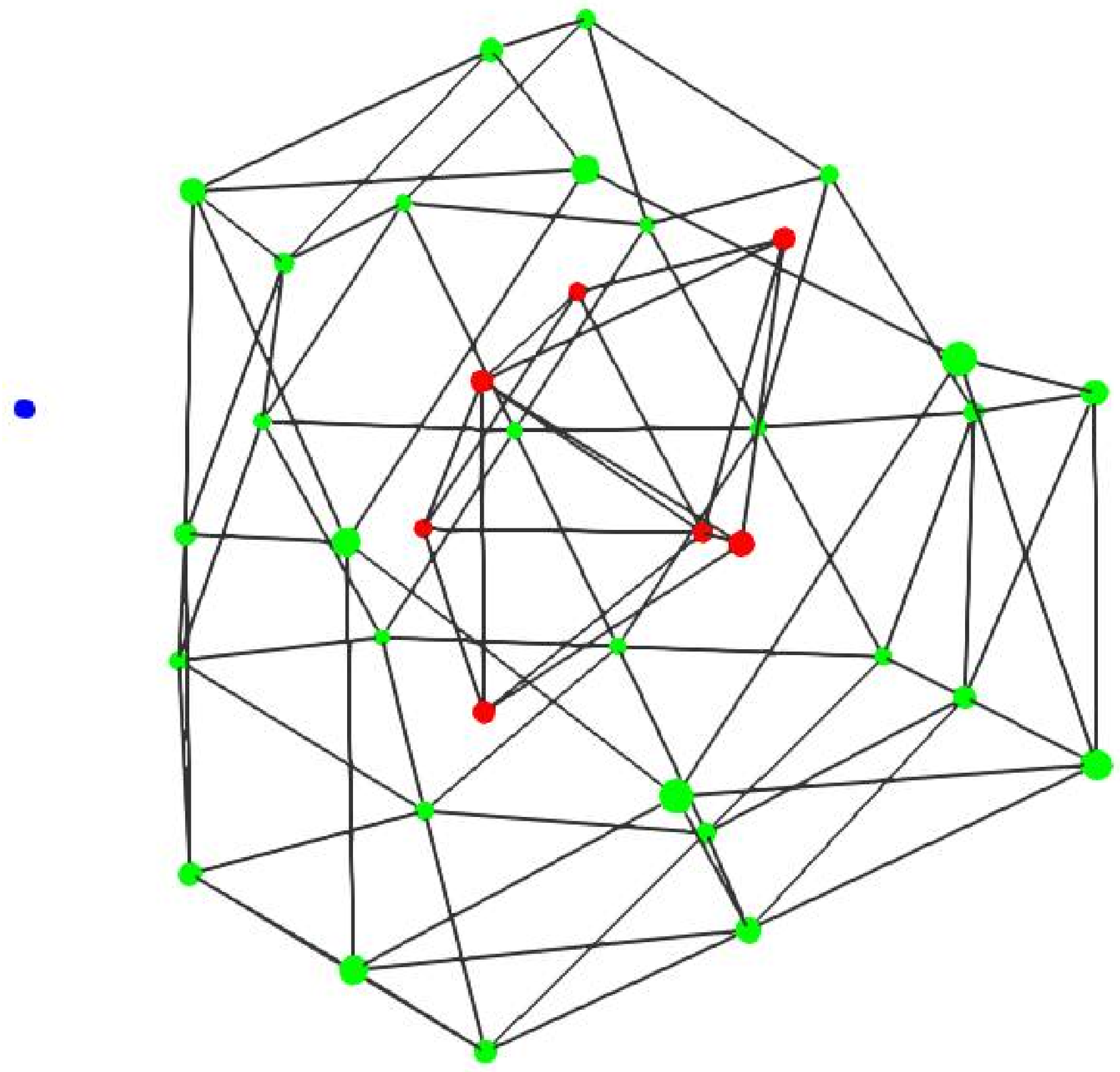, height=50mm}
\psfig{figure=\IMAGESPATH/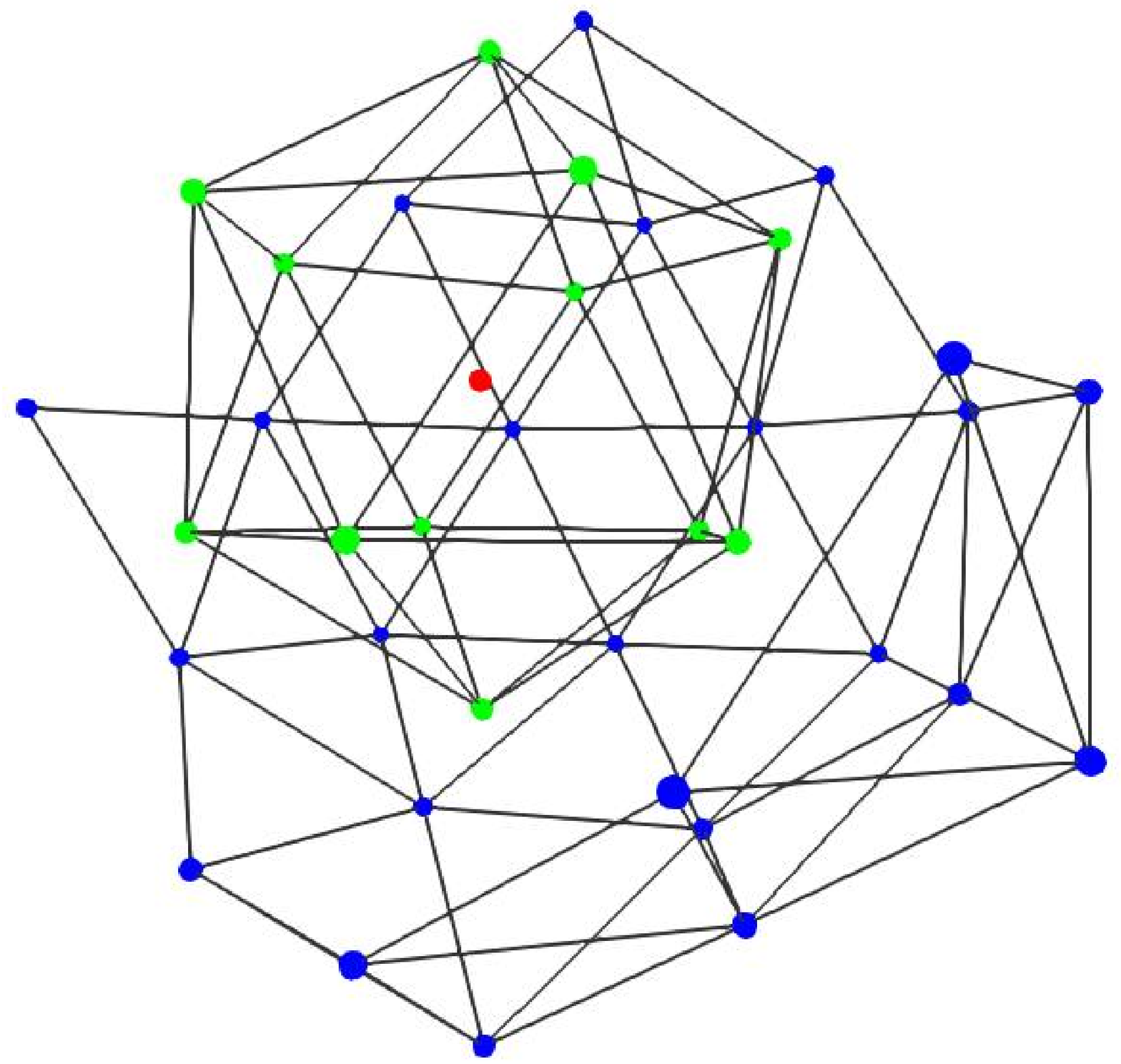, height=50mm}}
\centerline{ \makebox[2.2in][c]{ a) }\makebox[2.2in][c]{ b) } }
\caption{a) $C^*_{37}$'s view with nucleus n7 and b) $C^*_{37}$'s view with nucleus n1 IC.}~\label{fig:cl37}
\end{figure}

\begin{figure}
\centerline{ \psfig{figure=\IMAGESPATH/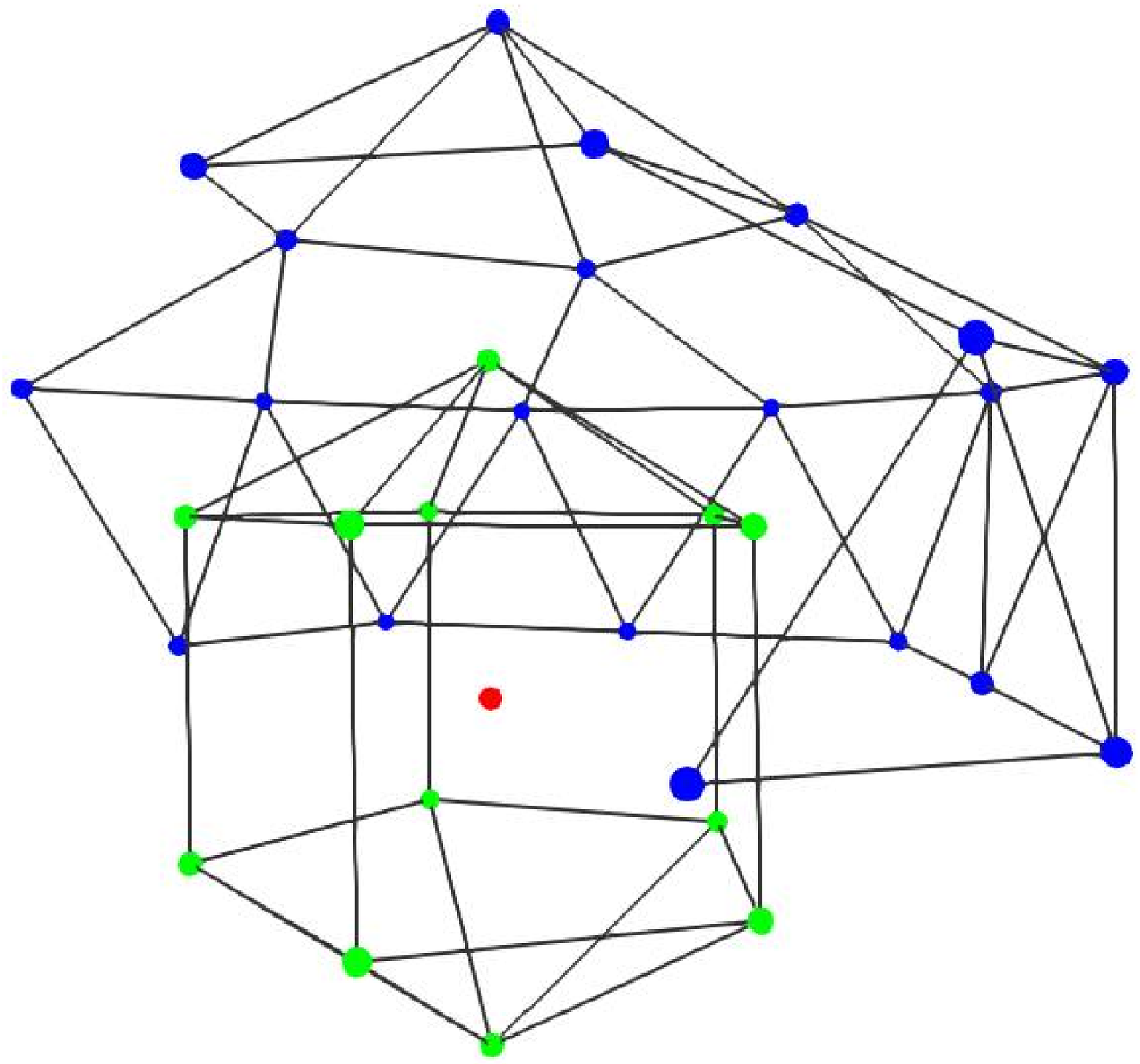,
height=50mm} \psfig{figure=\IMAGESPATH/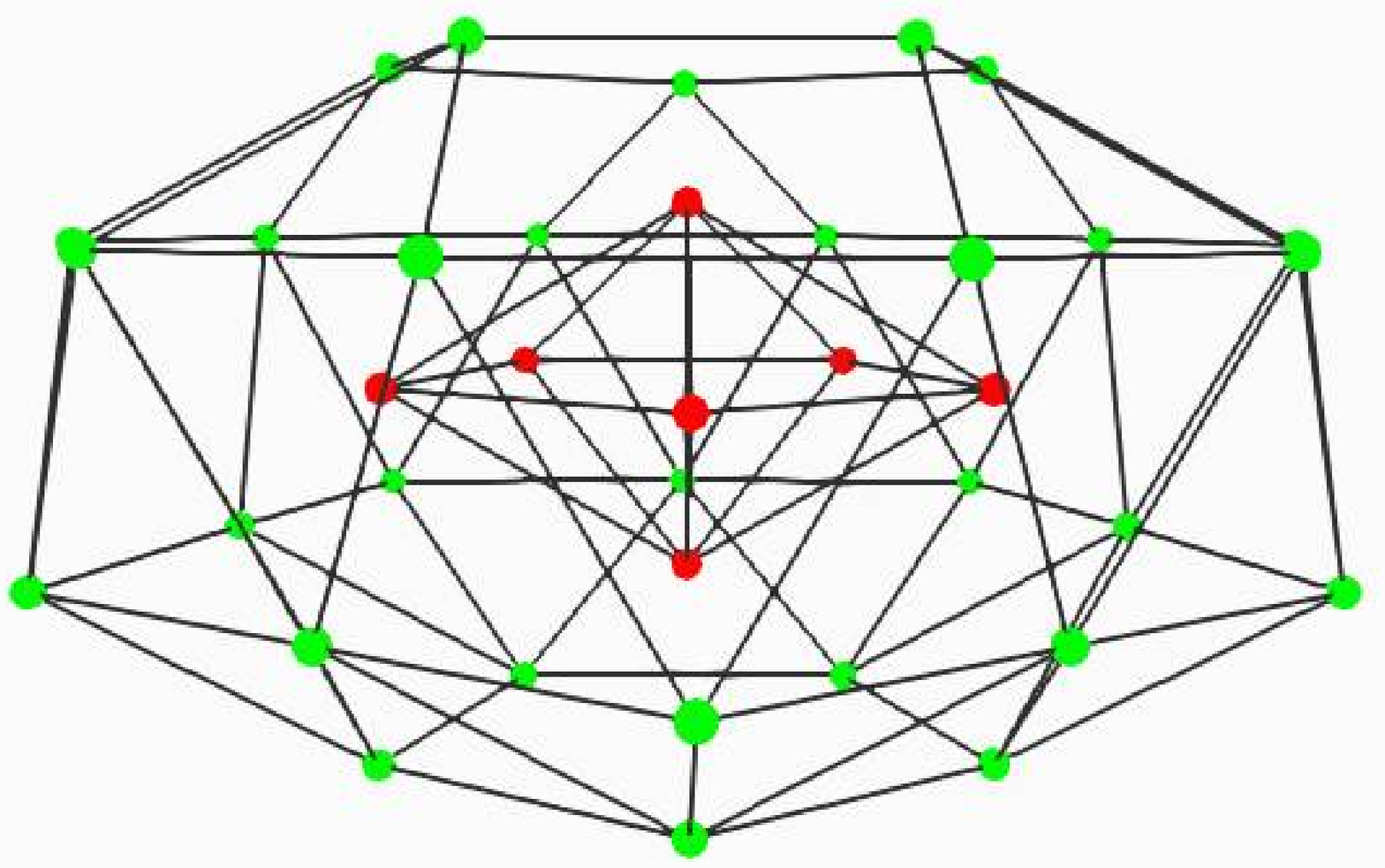,
height=50mm}} \centerline{ \makebox[2.2in][c]{ a)
}\makebox[2.2in][c]{ b) } } \caption{a) $C^*_{37}$'s view with nucleus n1 IR and b) non optimal, symmetric $C_{37}$ with nucleus n7 .}~\label{fig:cl37_2}
\end{figure}

\begin{figure}
\centerline{ \psfig{figure=\IMAGESPATH/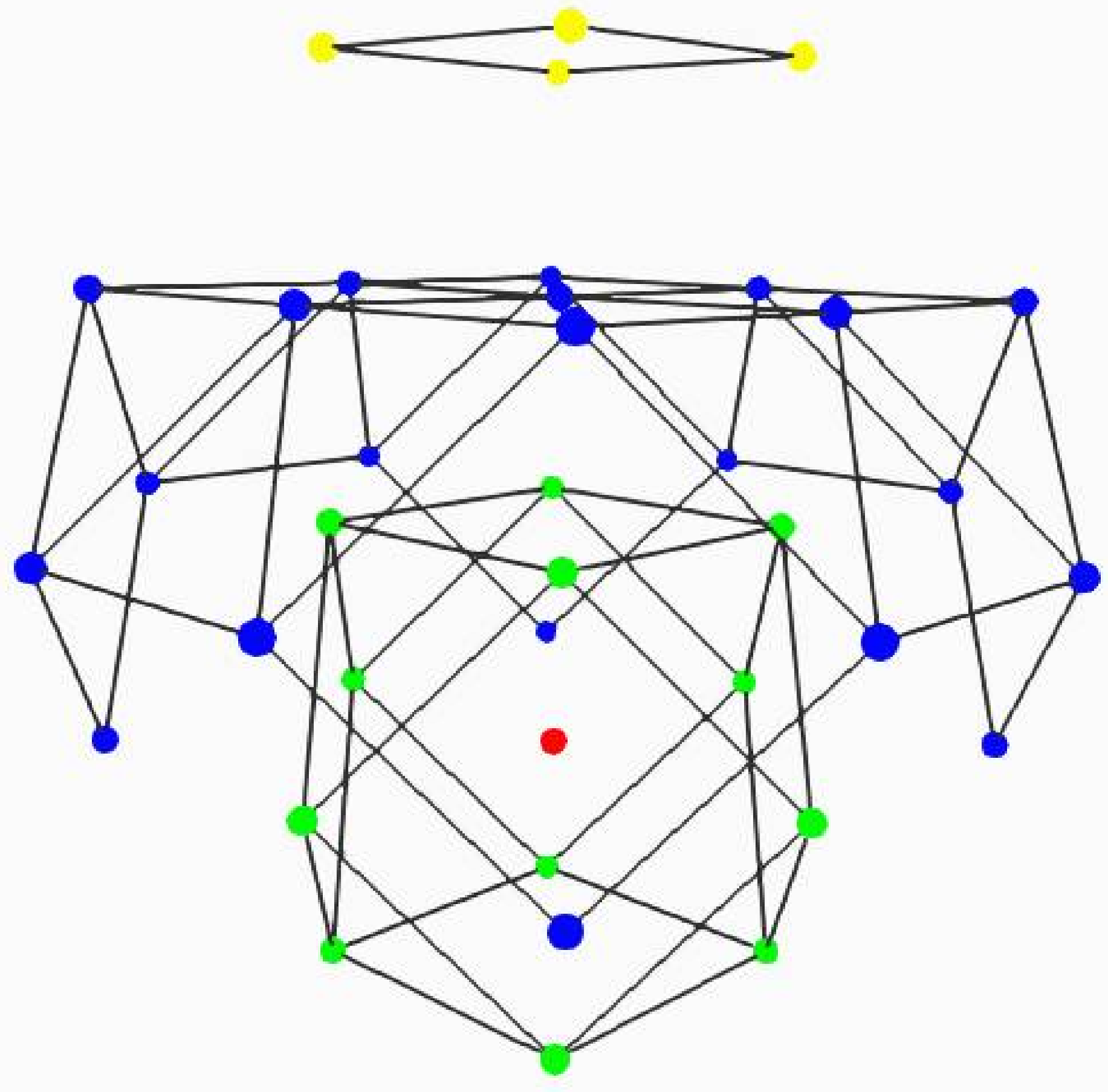, height=50mm}
\psfig{figure=\IMAGESPATH/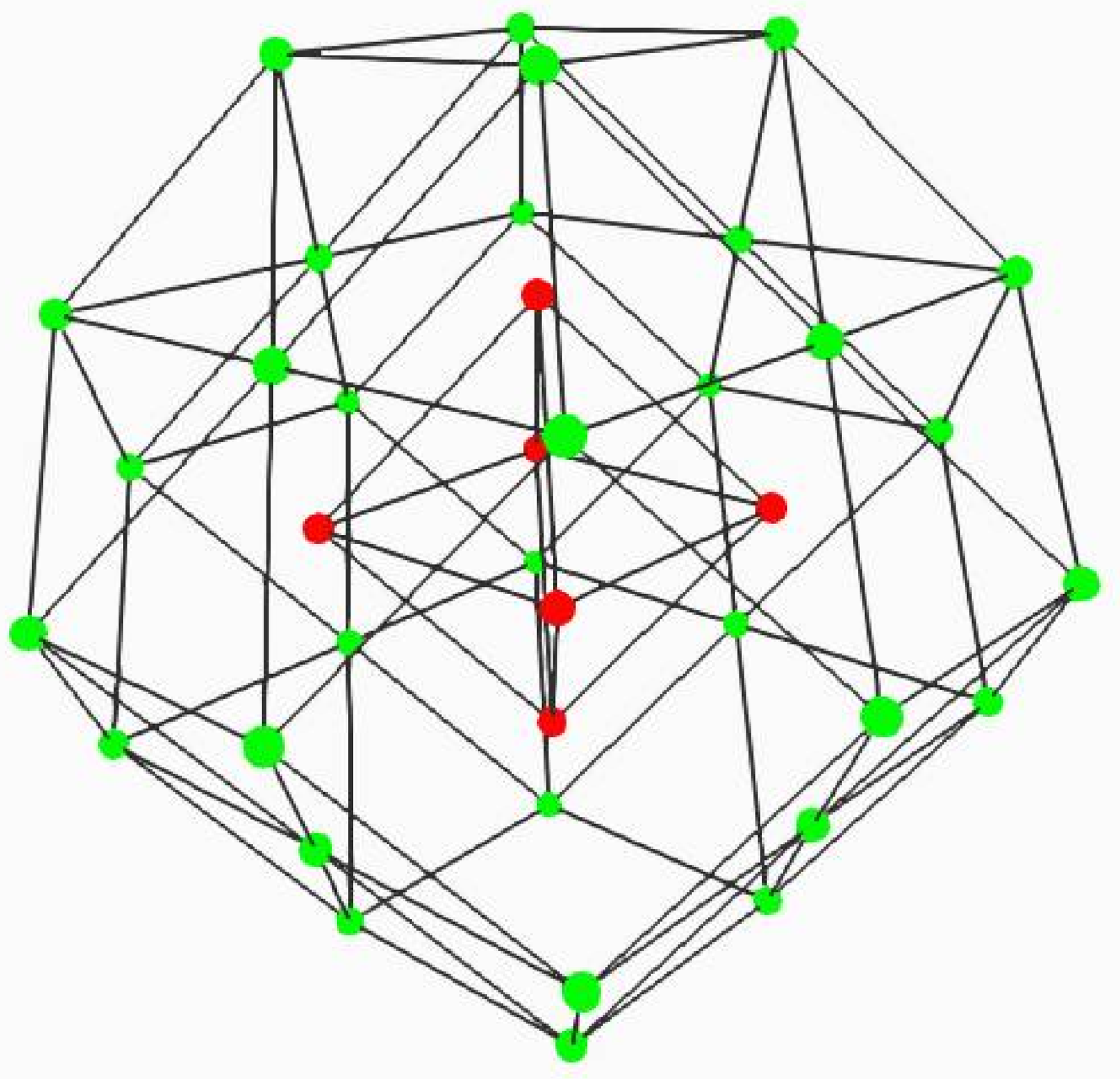, height=50mm}} \centerline{\makebox[2.2in][c]{ a) }\makebox[2.2in][c]{ b) } } \caption{
a) $C^*_{38}$'s view with nucleus n1, and b)
$C^*_{38}$'s classical view with nucleus n6}~\label{fig:cl38_n1_n6}
\end{figure}

\begin{figure}
\centerline{ \psfig{figure=\IMAGESPATH/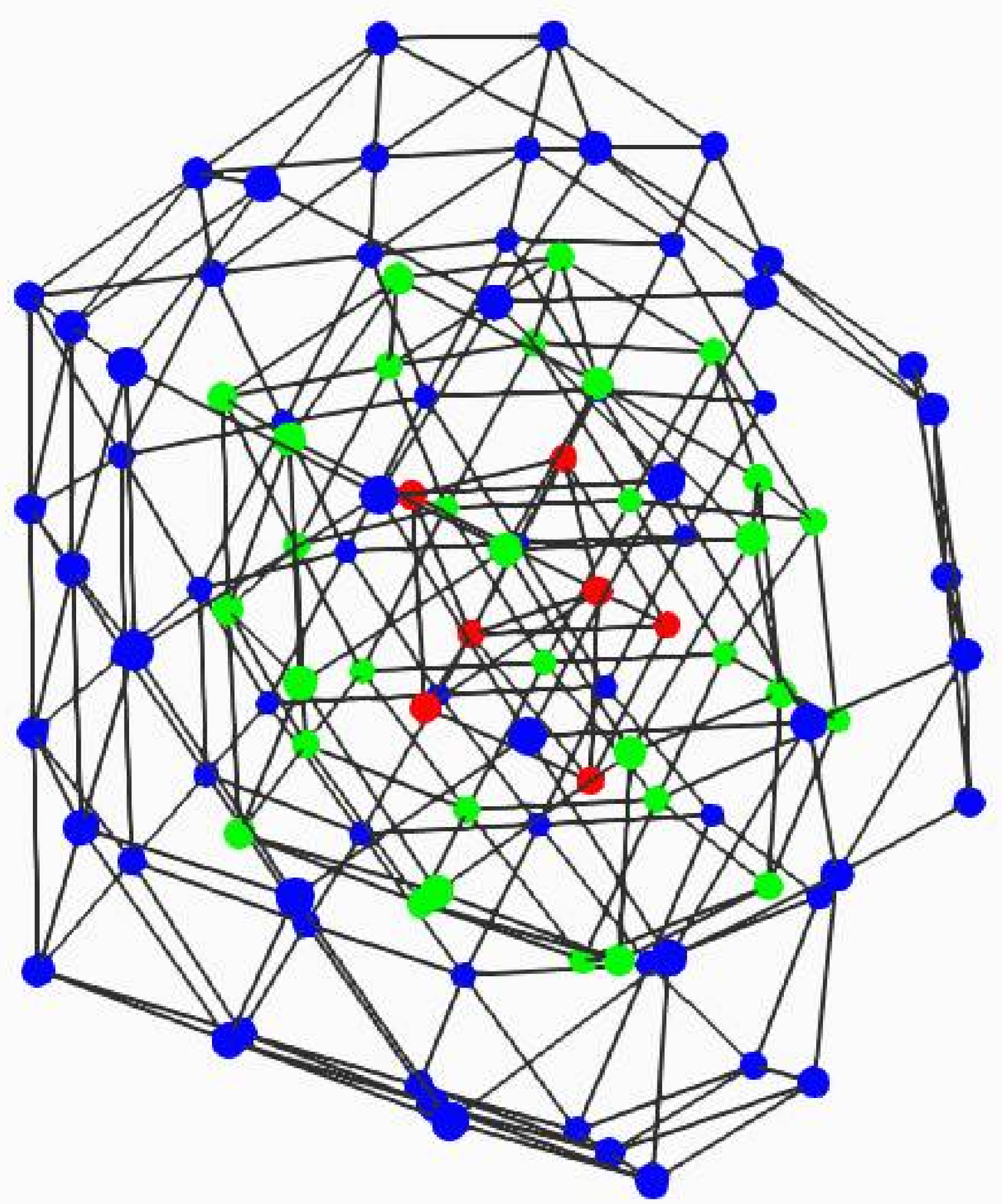,
height=50mm} \psfig{figure=\IMAGESPATH/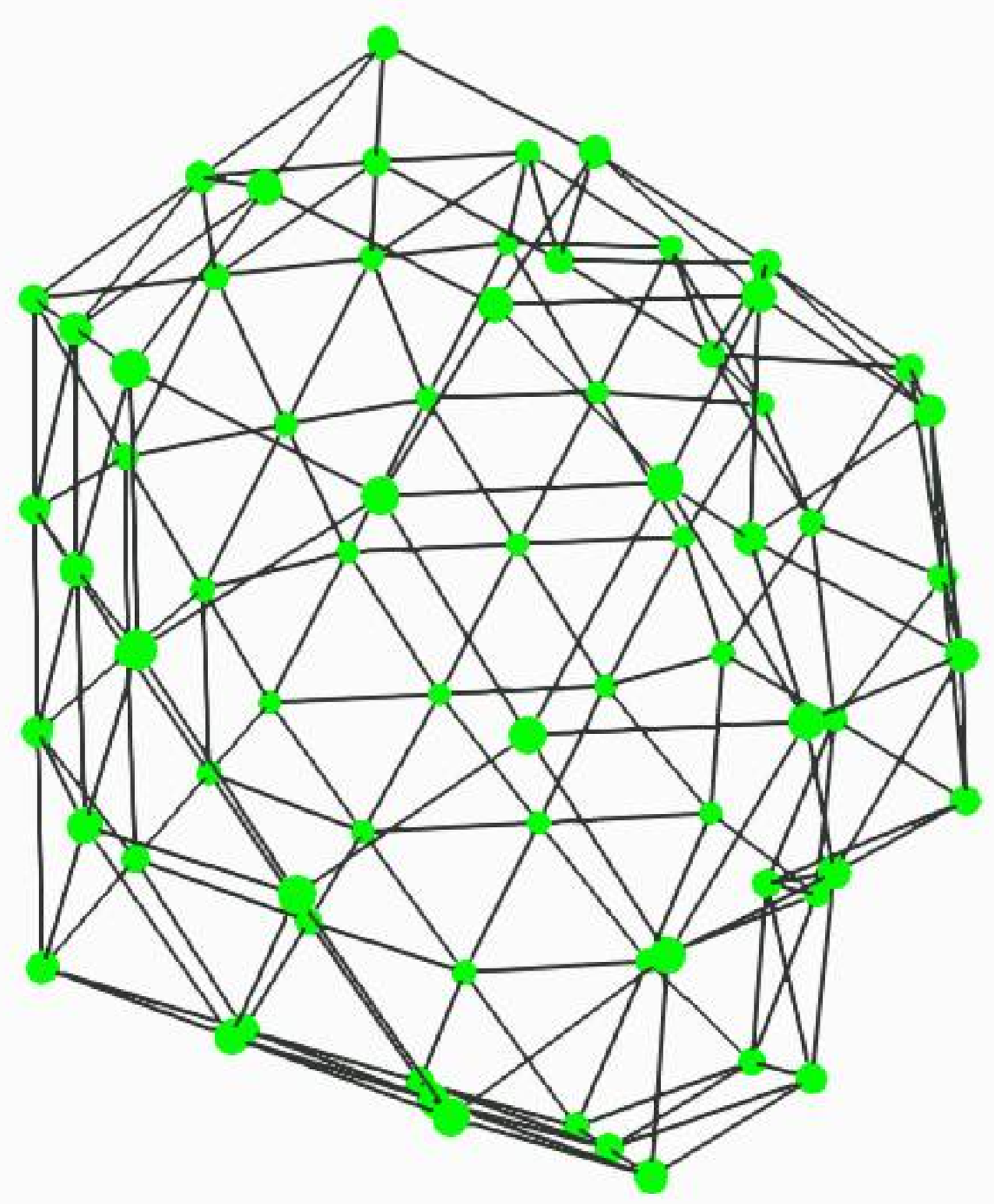,
height=50mm}}
\centerline{
\makebox[2.2in][c]{ a) }\makebox[2.2in][c]{ b) }}
 \caption{a) $C^*_{107}$ with nucleus n7, b) $C^*_{107}$'s shell}~\label{fig:cl107}
\end{figure}

\begin{figure}
\centerline{ \psfig{figure=\IMAGESPATH/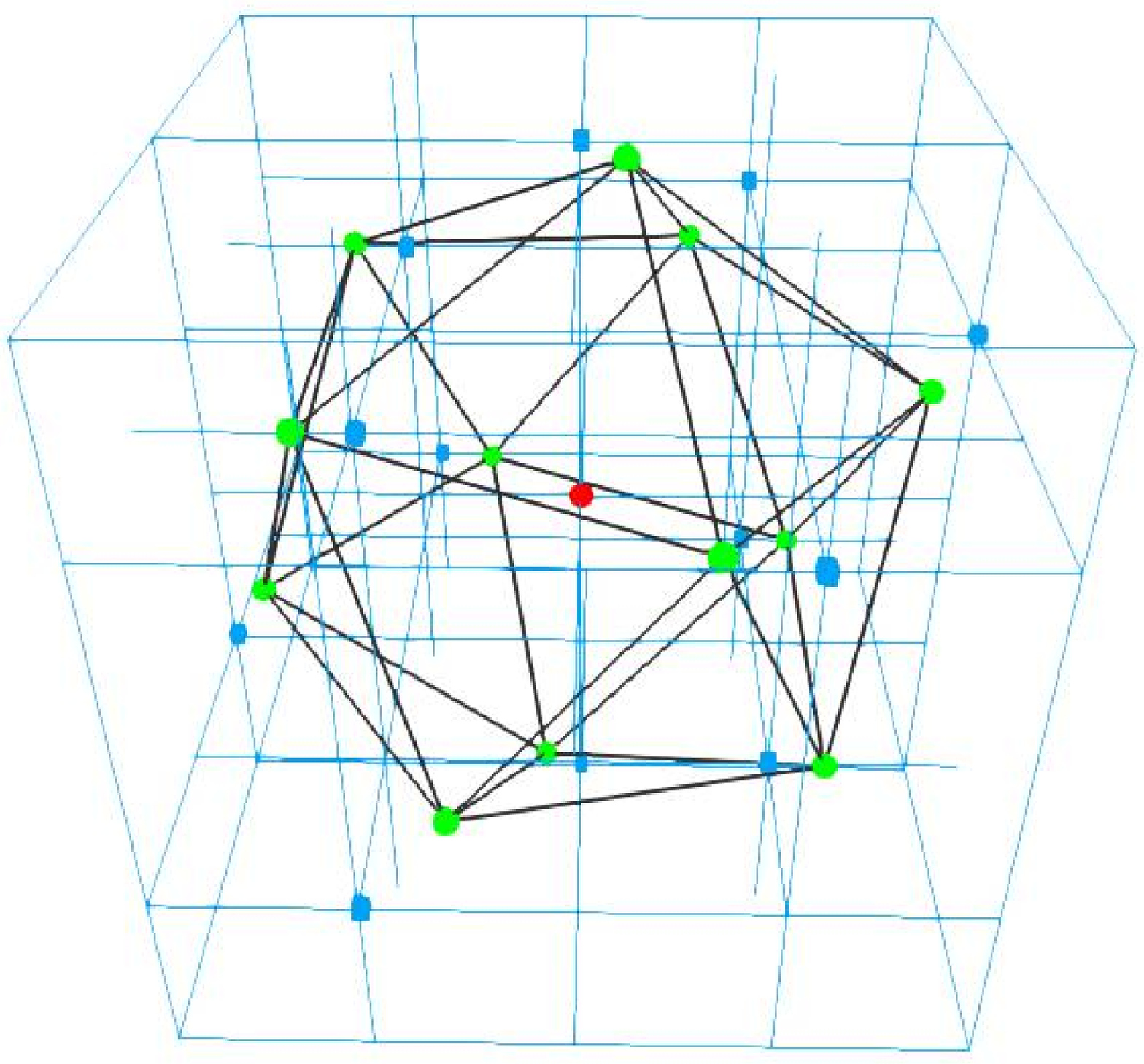, height=50mm}
\psfig{figure=\IMAGESPATH/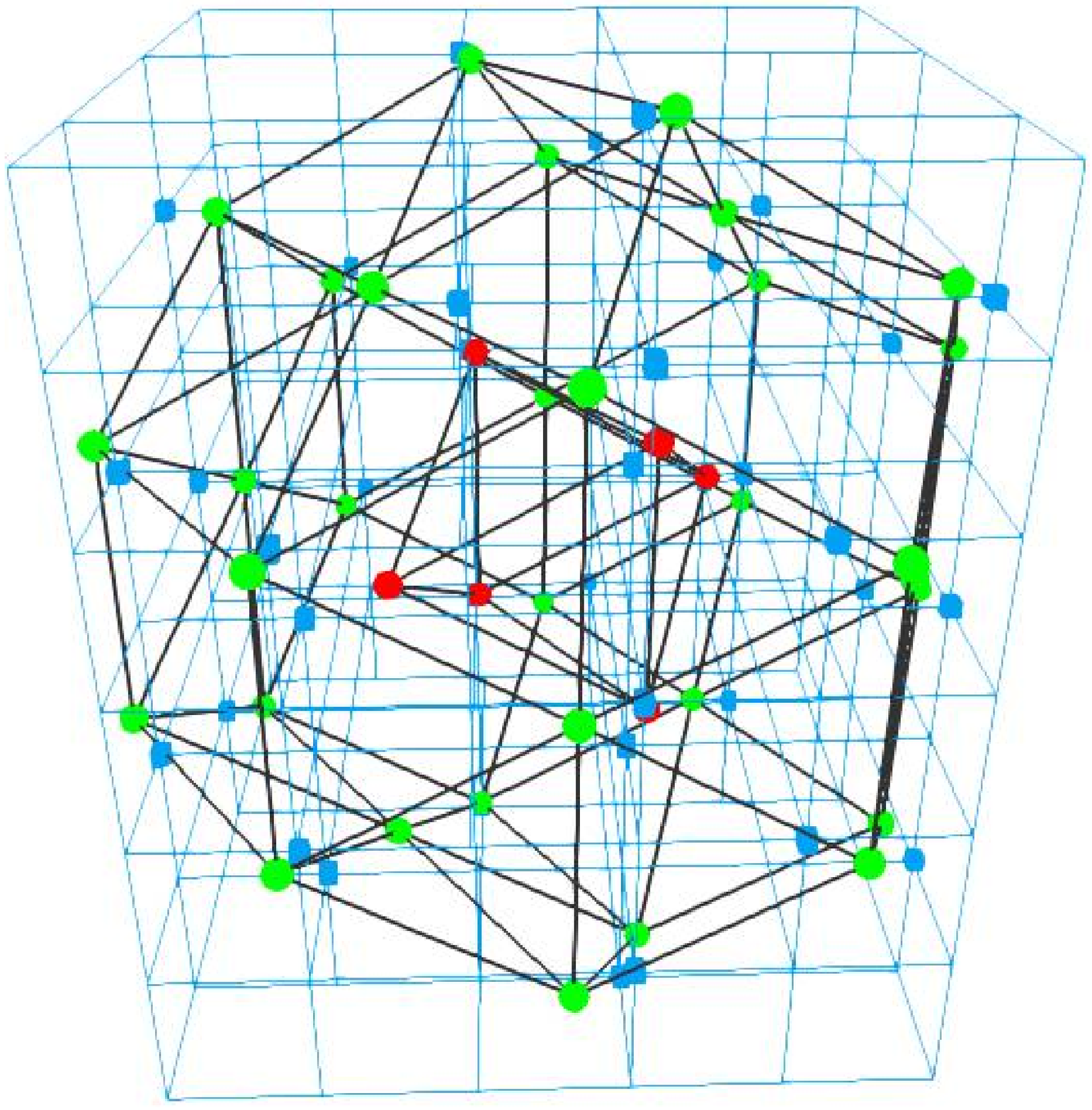, height=50mm}}
\centerline{
\makebox[2.2in][c]{ a) }\makebox[2.2in][c]{ b) }}
\caption{a) $C^*_{13}$ b) $C^*_{38}$ with their initial points inside of CB
lattice}~\label{fig:cl_13_38_inCB}
\end{figure}

Figure~\ref{fig:min_lattice1739} was constructed by the selection
of  not repeated positions from where a given initial selection of
particles converges always by a local minimization process to its putative minimal
LJ cluster. This set of positions is finite, and it can be
enumerate, such each position corresponds with a unique id number.

 Therefore, it could be simple to locate a minimal
LJ cluster by the set of its particles' number of the cluster. I proposed a telephone algorithm, which is like
make a phone call but, here $n$ id particles corresponds to a
cluster's phone number from a set of an appropriate selection of
$m$ id numbers from a region of IF lattice. After the minimization
if the value of the LJ potential is less than a previous cluster's
number, then it is the phone number of the cluster of $n$
particles. Even, it is like more genotype, I did not introduce
this type of DNA mechanics  in my previous genetic and evolutive
algorithms in order to keep a phenotype representation (this
means geometric shapes using the 3d particles' coordinate). The main idea was to look for the
putative optimal LJ cluster by an exhaustive searching. This is a
brute force algorithm with complexity related to the Newton binomio for 
combinations, $\binom{m}{n}.$

 Even with all
putative optimal LJ clusters from 2 to 1000 in the lattice IF, I
can not prove my conjecture. But, reviewing my previous work and
the mathematical properties of the LJ potential function, it is
possible to determine from the cubic lattice (CB) all optimal LJ
clusters in efficient time.

This paper presents an evolutive algorithm based in our previous
Genetic algorithm. It is based on the partial growing sequence
property that the optimal LJ clusters exhibe (To my knowledge, it
was Northby~\cite{jcp:Northby1987} the pioneer to state the
growing sequence property of the optimal LJ cluster over the IC
lattice, also Hoare~\cite{ap:Hoare1983} pointed out the 
morphology of the microclusters). It means that clusters with relative closed number of
particles could have similar geometry or in other words, they
belong to same lattice or they belong to the same geometrical family or
they shares some similar bricks or building blocks.

Some ideas and techniques are difficult to replicate, therefore
for this article, I added a simple Matlab programs to 
visualice my novel cluster partition and geometry, and to help for verifying my 8 categories of 
classification by similar number of particles considered the nucleus.

The corroboration of my results  was possible because all
the putative optimal LJ clusters are reported in The Cambridge Cluster Database (CCD)~\cite{http:CCD}.

The next subsection depicts the notation used. Section~\ref{sc:proper} has the properties and the proposition. The subsection~\ref{scc:partitionCL} depicts the technique for the creation of a partition of the cluster's particles into layers, and subsection~\ref{scc:heuristic} depicts an heuristic for determining a cluster's nucleus (in the appendix a Matlab program of such heuristic is depicted). Section~\ref{sc:parEvAlgorrithm} describes my version of parallel evolutionary algorithm. The next section presents the numerical results, and finally, the last section the conclusions and the future work.

%%%%%%%%%%%%%%%%%%%%%%%%%%%%%%%%%%%%%%%%%%%%%%%%%%%%%%%%%%%%%%%%%%%%%%%%%%%%%%%%%
%%%%%%%%%%%%%%%%%%%%%%%%%%%%%%%%%%%%%%%%%%%%%%%%%%%%%%%%%%%%%%%%%%%%%%%%%%%%%%%%%
%%%%%%%%%%%%%%%%%%%%%%%%%%%%%%%%%%%%%%%%%%%%%%%%%%%%%%%%%%%%%%%%%%%%%%%%%%%%%%%%%
\subsection{Notation}~\label{ssc:notation}
%%%%%%%%%%%%%%%%%%%%%%%%%%%%%%%%%%%%%%%%%%%%%%%%%%%%%%%%%%%%%%%%%%%%%%%%%%%%%%%%%
%%%%%%%%%%%%%%%%%%%%%%%%%%%%%%%%%%%%%%%%%%%%%%%%%%%%%%%%%%%%%%%%%%%%%%%%%%%%%%%%%
%%%%%%%%%%%%%%%%%%%%%%%%%%%%%%%%%%%%%%%%%%%%%%%%%%%%%%%%%%%%%%%%%%%%%%%%%%%%%%%%%
Given a set $S$, $|S|$ is
the number of elements of the set. Also if $A[\cdot]$ is an array,
$|A[\cdot]|$ is the number of elements of the array. $\emptyset$
is the empty set. $||\cdot||$ is the norm in $\Real^3$.

Particularly, $C^*_{n}$ denotes an optimal LJ cluster with $n$ particles, and 
$C_{n}$ denotes an arbitrary cluster with $n$ particles.

A cluster $C_n$ or $C^*_n$ are sets of natural numbers, where
each number correspond to a particle's properties ($p_i$). 
For this research the particle's properties are 
the particle's 3D coordinates. $|| p_i,p_j ||$ is the Euclidian distance
 between particles $p_i$ and $p_j$.
 
By example,  $C^*_n=\{1,2\}$, $p_1=(-\frac{d^*}{2},0,0)$, and
$p_2=(\frac{d^*}{2},0,0)$ where $d^*=\sqrt[6]{2}$ 
is the optimal distance for two particles under LJ potential:
$$LJ(d) = d^{-12} - 2 d^{-6}$$

Several references explain how to build IC and FC
lattices~\cite{jgo:Leary1997, ps:Maier1992, ape:Solovyov2003,
jpca:Xiang2004A}. The CB lattice is very simply is the set of
points that correspond to the intersection of the parallel lines
to the axes with a separation of $d^* /2$ from the (0,0,0).

%%%%%%%%%%%%%%%%%%%%%%%%%%%%%%%%%%%%%%%%%%%%%%%%%%%%%%%%%%%%%%%%%%%%%%%%%%%%%%%%%
%%%%%%%%%%%%%%%%%%%%%%%%%%%%%%%%%%%%%%%%%%%%%%%%%%%%%%%%%%%%%%%%%%%%%%%%%%%%%%%%%
%%%%%%%%%%%%%%%%%%%%%%%%%%%%%%%%%%%%%%%%%%%%%%%%%%%%%%%%%%%%%%%%%%%%%%%%%%%%%%%%%
\section{Properties of LJ}~\label{sc:proper}
%%%%%%%%%%%%%%%%%%%%%%%%%%%%%%%%%%%%%%%%%%%%%%%%%%%%%%%%%%%%%%%%%%%%%%%%%%%%%%%%%
%%%%%%%%%%%%%%%%%%%%%%%%%%%%%%%%%%%%%%%%%%%%%%%%%%%%%%%%%%%%%%%%%%%%%%%%%%%%%%%%%
%%%%%%%%%%%%%%%%%%%%%%%%%%%%%%%%%%%%%%%%%%%%%%%%%%%%%%%%%%%%%%%%%%%%%%%%%%%%%%%%%

There are several articles about LJ potential function's
properties. The proposition 1 in~\cite{arXiv:Barron2005} is
repeated as proposition 2.1 in~\cite{arXiv:Barron2010}, together
with proposition 2.2:

{\bf Proposition 2.1} Exist a discrete set, $\Omega $, where
$\forall j\in N$, $j\geq 2$, the potential of SOCDXX($j$) has the
same ("close value") optimal value of SOCCXX($j$) for a potential function such
that

\begin{enumerate}
    \addtolength{\topsep}{-\baselineskip}
    \addtolength{\itemsep}{-\baselineskip}
    \setlength{\parskip}{\baselineskip}
\item  $\lim_{r_{i,j}\rightarrow 0}\hbox{VXX}\left( r_{i,j}\right)
=\infty$. \item  $\nabla ^{2}\hbox{VXX}\left( x^{\ast }\right) $
semi-positive$,\left\| \nabla \hbox{VXX}\left( x^{\ast }\right)
\right\| \ll 1$ and $\frac{\left\| \nabla \hbox{VXX}\left( x^{\ast
}\right) \right\| }{\left| \hbox{VXX}\left( x^{\ast }\right)
\right| }<\delta _{0}$, where $0<\delta _{0}\ll 1$
\end{enumerate}
where XX is BU or LJ.

where SOCYXX means search for optimal cluster, D is discrete, C is
continues and XX is LJ for Lennard Jones Potential or BU for
Buckingham potential.

{\bf Proposition 2.2} Any shape of $n$ particles with edges
$\approx d^\ast$ can be approximated from the CB lattice.

This means that with an appropriate region of CB lattice is
sufficient to look for optimal clusters of size $n$. I did not state the size of  the appropriate region of CB. However, today, any optimal LJ cluster in the CCD has an initial configuration in CB lattice, such that from this initial configuration converges by a minimization process to its corresponding putative optimal LJ cluster.

\begin{proposition}
~\label{prop:CBUpper} For any set of particles of a cluster's $CL$ or 
a set of particles of a region $RL$ of any lattice based on a unit $u$. 
Then corresponding region $RB$ of the CB lattice such 
$RB$ covers them under the $||\cdot||_\infty.$
 Then particles of $CL$ or $RL$ are less than the number of particles of $RB$, i.e.,
 $|CL| < |RB|$ or $|RL| < |RB|$, 
where $|\cdot|$ is the number of particles.
\begin{proof}
Under the $||\cdot||_\infty$ any region of CB is a 3D cube. By construction, 
it has a point at the center $(0,0,0)$, the first cube with 
$-u/2,0,u/2$ has $3^3$ points, 
the second cube with $-u,-u/2,0,u/2,u$ has $5^3$ points, $\ldots,$ 
the $k$ cube has $(2k+1)^3$ points.
Any polyhedra or lattice based in the unit $u$ 
can not have more than 12 neighbors at ratio $u$. 
The icosahedra has 13 points but the corresponding cube to cover it is the second cube,
with $5^3$ points, i.e., no. particles of icosahedra < no. particles of unit cube of CB lattice. 
 In general,  for a given cluster $CL$ or region $RL$ they can be divided and contained 
by a set of unit cubes of $CB$, which is a cube, let's call $RB$. 
Therefore, $|CL| < |RB|$ or $|RL| < |RB|$.
\end{proof}
\end{proposition}

It follows that any region of the CB lattice has more points than the same region of a lattice.

But more important,  the global 
continuos optimal LJ cluster can be approximated in a discrete set of 3D points, $\Omega$. 
Then a connection between $RB$ an appropriate region and $\Omega$ will be provide a discrete
set of 3D points where the continuous optimal global cluster is 
approximated by a discrete set of points! 
A local minimization procedure is the connection to approximate the continuous optimal global in 
$RB$. 
On the other hand, for a cluster with $n$ particles, let's suppose to have an appropriate 
region $RB$, $m=|RB|.$ 
The number of the posible clusters of size $n$ in $RB$ is $\emph{M} = \binom{n}{m} \gg 0.$

Note that $\emph{M}$ is a big number. It follows naturally from prop.~\ref{prop:CBUpper} 
than for any
region of a given lattice, the number of clusters with $n$ particles is $\ll$  
$\emph{M}.$

\begin{proposition}
~\label{prop:ProbCB_Lats} $RB$ is an appropriate search  region of the CB lattice for
a cluster with $n$ particles, $\emph{M} = \binom{n}{m}, m=|RB|,$ $\emph{M}$ is a 
huge positive number. $|RL|$ is a region of a lattice 
where there are different clusters with $n$ particles, 
and it is supposed that it contains the optimal LJ cluster.
Then 
$$\textsl{P}\left(\frac{F_1}{RB}\right) < \textsl{P}\left(\frac{F_2}{RB \cup RL}\right)$$
where $\textsl{P}(\cdot)$ is a probability function, $F_1$ is the set of the
optimal candidates for being the global optimal LJ cluster in $RB$, and
$F_2$ is the set of the
optimal candidates for being the global optimal LJ cluster in $RB \cup RL.$  
 
\begin{proof}
It follows from
$$ |RL| f_1 < f' \emph{M}$$ 

where $f_1=|F_1|$, $f'>0$ is the number of candidates for being the global optimal LJ cluster in 
$RL$,  
$f'>0,$ it is not zero because the assumption that $RL$ contains the optimal LJ cluster,
$f_1 \ll \sqrt{\emph{M}}$, and  $|RL| \ll \sqrt{\emph{M}}$.
Then
$$ (\emph{M}+|RL|) f_1 < (f_1+f')\emph{M}, $$
$$ \textsl{P}\left(\frac{F_1}{RB}\right)=\frac{f_1}{\emph{M}} < 
 \frac{f_1+f'}{\emph{M}+|RL|} = \textsl{P}\left(\frac{F_2}{RB \cup RL}\right)$$
\end{proof}
\end{proposition}

It is important to assume that $F_2 \cap RL \neq \emptyset$, to increase the
probability for determining the global optimal clusters, otherwise the probability does not increase. Many of the ad-hoc, heuristic,
genetic, and evolutionary algorithms for determining the optimal LJ clusters have been 
used this property as previous knowledge to favorece some candidates over others with 
success and speed to replicate the putative optimal LJ clusters. 

In a personal communication, I suggested at 2004 to Shao, et al. to use 
different lattices from~\cite{ape:Solovyov2003}. In~\cite{pc:Shao2005} 
appears the acknowledge: "The authors would like to thank Prof. 
Carlos Barr\'{o}n Romero for his personal
communications and collaborations with us in the studies on the lattice-based
optimization methods, including also the work published in J. Phys. Chem. A, 108,
3586-3592 (2004)."

So even, knowing that an appropriate region of the CB lattice has the optimal LJ clusters, to
improve the efficiency for determining  optimal LJ cluster is a good strategy to
use other sources of candidates to favorece diversity in the complex process 
for looking  the unknown optimal LJ clusters.  

Finally, the theoretical results point out that it is posible to increase the speed
of any algorithm for determining optimal LJ clusters but without any 
proof that they are global optimal. The repeated putative optimal LJ clusters are  
stationary states, from where a criteria such of the number of times that the same cluster 
appears, then stop and accept it as the putative optimal global LJ cluster. 

The number of steps in these cases are clearly very less than $\emph{M}=\binom{n}{m}, m = |RB|$.
$\emph{M}$ is the huge number related to the numbers of candidates to compare for 
determining global optimality in $RB$.
A force brute algorithm for the estimation of the different combinations
of a set can be found in~\cite{arXiv:Barron2010}, it is a version for 
determining the different cycles of a complete graph, $G=(V,A), |V|=n$.

%========================================================
%========================================================
\subsection{Partition technique for a cluster}~\label{scc:partitionCL}
%========================================================
%========================================================

The geometry of the LJ clusters have been strongly related to different geometric structures (see~\cite{ap:Hoare1975, ap:Hoare1983, jcp:Northby1987, ape:Solovyov2003}) icosahedral, dodecahedral, cuboctahedral, and so on. 

My segmentation's technique provides different cluster's views as an arbitrary polyhedron with its partitioning into its core, layers and shell by using the particle's neighbors. The advantages to segment a cluster with my technique are 1) to help for interpretation and interaction with other clusters and lattices, and 2) to build a cluster from lego or building blocks.

These properties are quite important because they support the previous research 
about the knowledge 
of the clusters' morphology, properties, geometrical families, chemistry, 
or the well know grow sequence.

A particle's neighbor structure is defined as follow:
\begin{enumerate}
    \item Define a unit: $u$.
      \item Define a tolerance $t$ as the porcentaje for accepting the expansion and       the compression of $u.$ ($0 \leq t \leq 1.0$). 
      \item Neighbor's criteria: Particles $p_i,p_j$ are neighbor
       if an only if $(1-t) u < dist (p_i -p_j) < (1+t)        u.$ 
       \item for each particle, $Nvec(i)$ is the        number of neighbor of the particle $p_i$,       and $Vec(i,k)$ is        the array for storing the number of the particles $p_k$ that 
       they satisfy the neighbor's criteria with a given particle $p_i$.   
\end{enumerate}    

where $dist$ is an appropriate distance function between the particles. 
$p_i$ stands for particle's representation in a n dimensional space,
a particle is represented by $p_i$, 
which it could contain all the relevant particle's attributes. 

For this research, $u=d^*$,  
$p_i$ is the particle's 3D coordinates, $dist = || \cdot||$ is the Euclidian distance, 
and $t=0.1$.

One characteristic of LJ clusters is the compression-expansion over the 
distance between particles with respect to the unit $d^*$. 
The value of $t=0.1$ allow to differentiate a diagonal from a expanded-compressed unit $u$
 and it works well
 to identify the "hard LJ optimal clusters" 
 (see ~\cite{HartkeDittner2016}, by example clusters with 38, 75, 98,75, 76, 77, 102, 
 and 103 particles). For any possible LJ cluster, the upper limit of 12 neighbors 
 over its particles, i.e.,  
 $Nvec(i) \leq 12, \forall i.$ comes from the upper limit inherited by 
 the 3D twelve kissing spheres geometrical property.

With the cluster's neighbors information, 
the next algorithm builds an arbitrary  partition of a cluster's particles into 
a set of layers: 

\begin{algorithm}~\label{alg:partCluster} Partitioning a cluster $C_n$

\noindent\textbf{input:} $C_n$: array of int, for the set of particles' numbers;

$Nvec$: array of int, with the particles' number of neighbor;

$Vec[i,k]$ : array of [int, int] with the particles' neighbors;

$Nuc=\{i_0,i_1,\ldots,i_k\}$: array of int, for a given set of particles's number to be the nucleus, 
with $|Nuc| \leq n$;

\noindent \textbf{output:} $capa$: array of int, for  the corresponding layer of a particle;

$ncapa$: int, for the cluster's number of layers;

\noindent \textbf{memory}: $fmk$: int;

%\vspace{-1mm}
\noindent\hrulefill
\begin{easylist}
\renewcommand{\labelitemi}{\ }
\setlength\itemsep{-0.05cm}

\item {\bf for} i := 1 {\bf to} $n$ {\bf do}
\item \hspace{0.5cm}  $capa[i]:=0$;
\item {\bf end for}

\item $ncapa := 1$;
 
\item {\bf for} i := 1 {\bf to} $|Nuc|$  {\bf do}
\item \hspace{0.5cm} $capa[Nuc[i]]: = ncapa$
\item {\bf end for}

\item $fmk := 0$;
\item {\bf while} (1) {\bf do} 
\item \hspace{0.5cm} {\bf for} i:=1 {\bf to} n  {\bf do}

\item \hspace{0.5cm}\hspace{0.5cm}{\bf if} ($capa[i] == ncapa$) {\bf then} 
\item \hspace{0.5cm}\hspace{0.5cm}{\bf for} $jv := 1$ {\bf to} $Nvec[i]$  {\bf do} 
\item \hspace{0.5cm}\hspace{0.5cm}\hspace{0.5cm}$kv := Vec[i][jv]$;
\item \hspace{0.5cm}\hspace{0.5cm}\hspace{0.5cm}{\bf if} ($capa[kv] == 0$) {\bf then} 

\item \hspace{0.5cm}\hspace{0.5cm}\hspace{0.5cm}\hspace{0.5cm}$capa[kv] := ncapa$ + 1;

\item \hspace{0.5cm}\hspace{0.5cm}\hspace{0.5cm}\hspace{0.5cm} $fmk := 1$;

\item \hspace{0.5cm}\hspace{0.5cm}\hspace{0.5cm}{\bf end if}
\item \hspace{0.5cm}\hspace{0.5cm}{\bf end for}

\item \hspace{0.5cm} $ncapa := ncapa$ + 1;

\item {\bf end while}
\item  $ncapa := ncapa$ - 1;

\item \textbf{return};

\end{easylist}
%\vspace{-2mm}
\noindent\hrulefill
\end{algorithm}

Hereafter, layer number 1 is the core or nucleus, and the last layer $ncapa$ is the shell.

It is easy to verify that the set $capa$ of the  
particles' numbers is a partition of $C_n$, i.e,:
\begin{itemize}
    \item $\bigcup_{i=1}^{ncapa} capa[i]$ = $C_n$
    \item $capa[i] \cap capa[j] = \emptyset$, $i \neq j.$
\end{itemize}     

\begin{figure}
\centerline{ \psfig{figure=\IMAGESPATH/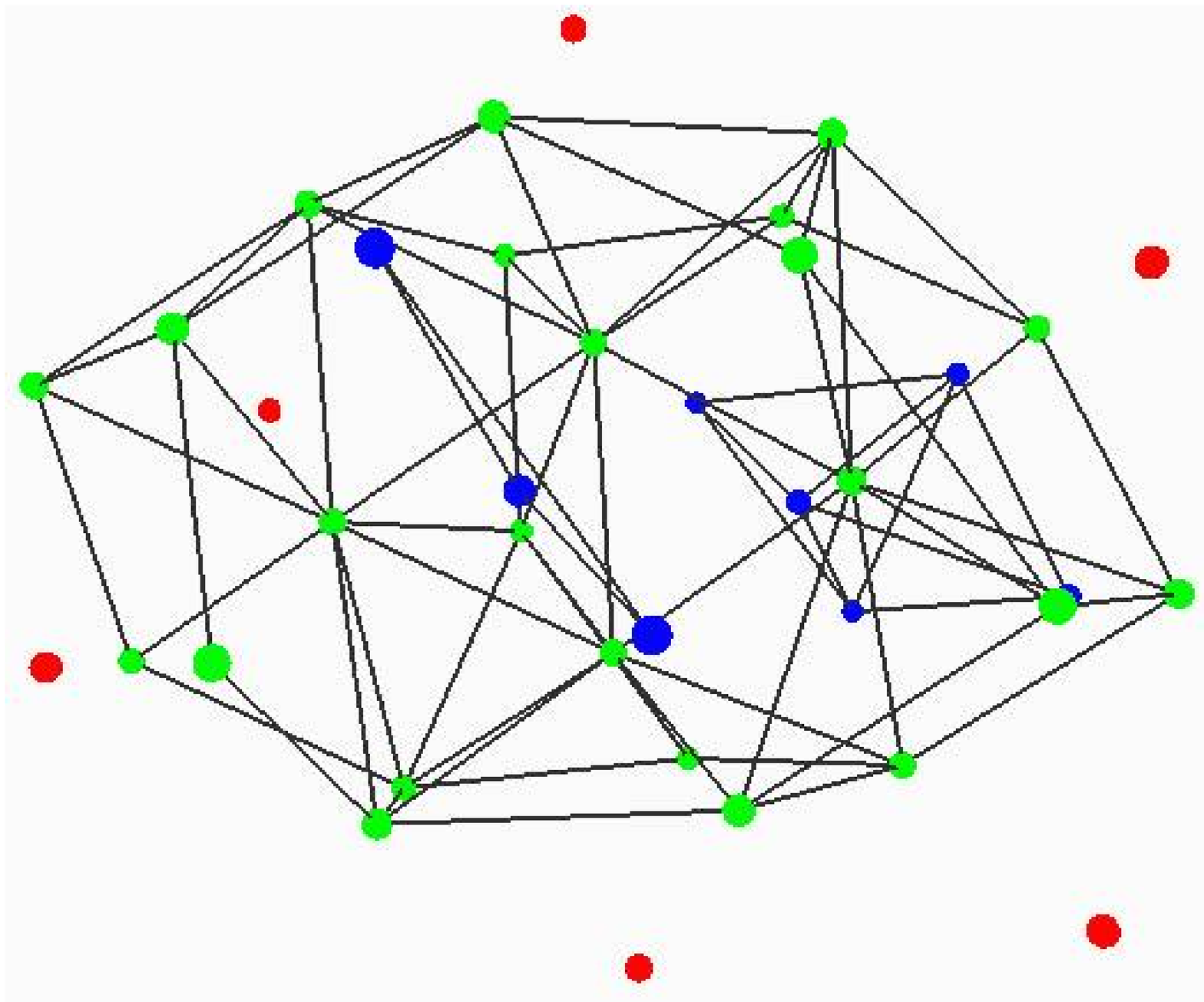,
width=40mm}
\psfig{figure=\IMAGESPATH/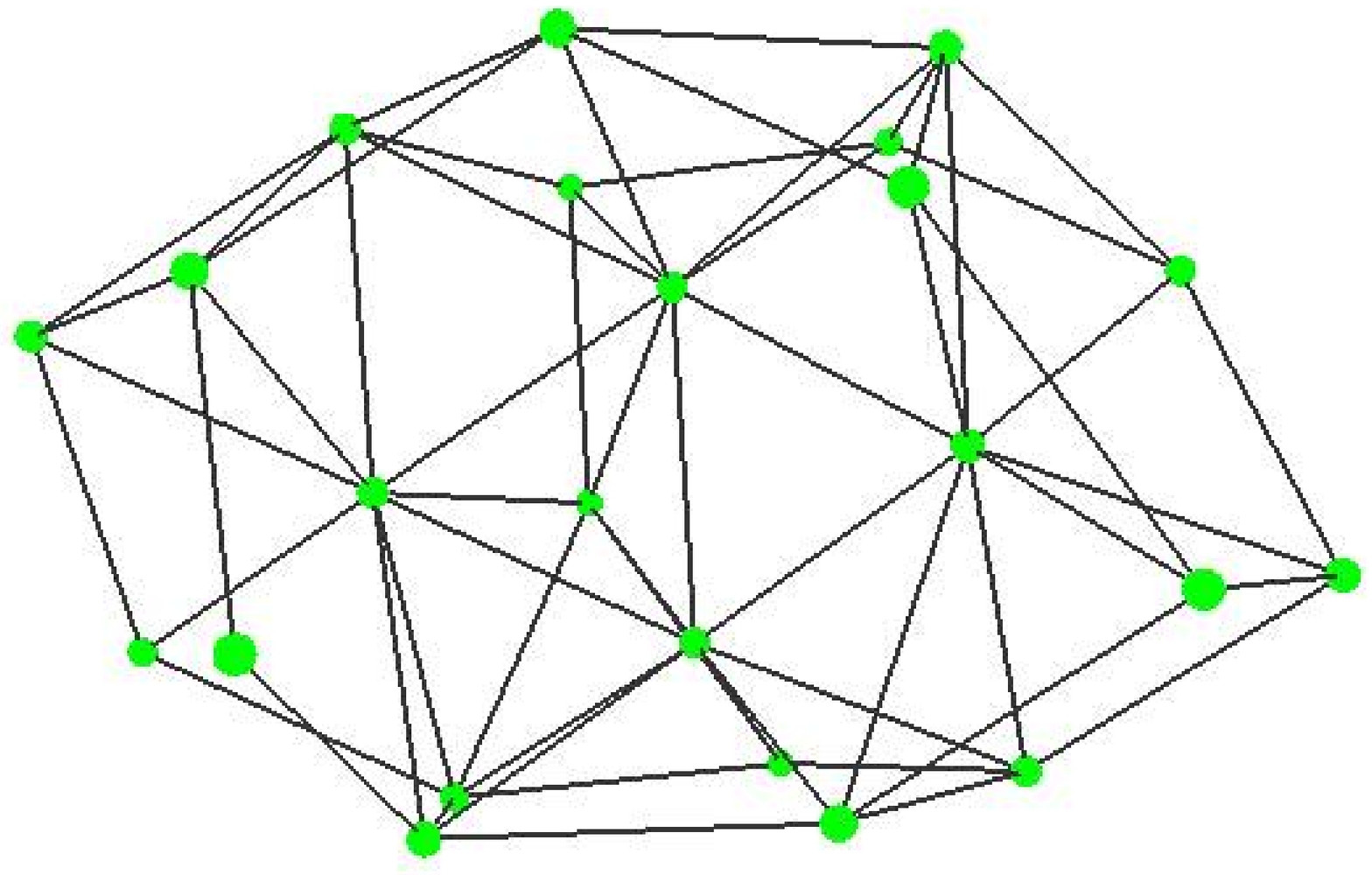,
width=40mm}
\psfig{figure=\IMAGESPATH/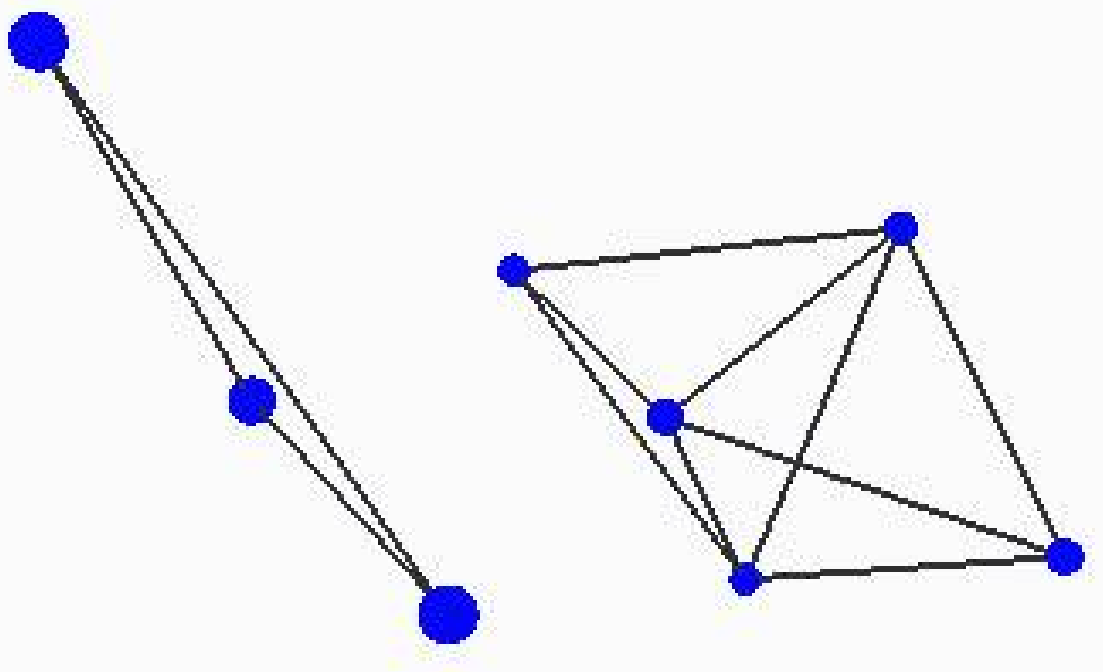,
width=40mm}} 

\caption{$C^*_{37}$ with its layers for  an arbitrary selection of its nucleus}~\label{fig:cl37NucArb}
\end{figure}

The previous algorithm gives an arbitrary cluster partition that it could not correspond 
to the standard accepted geometric structures but it can be used for genetic 
cuts for creating a new offspring as playing with set of figures of lego. 
In particular, the results could be similar to the Hoare's (see \cite{ ap:Hoare1983}) 
morphology of simple microclusters, 
polyhedral structures (PT), and an arbitrary representation of cluster's isomeros. Figure 
~\ref{fig:cl37NucArb} 
depicts $C^*_{37}$ with its layers for  an arbitrary selection of its nucleus. Figure~\ref{fig:cl107} depicts $C^*_{107}$ with nucleus n7 and its shell.

%========================================================
%========================================================
\subsection{Heuristic for determining a nucleus 
for a cluster}~\label{scc:heuristic}
%========================================================
%========================================================

The proposed heuristic is simple and it is based in previous knowledge of the 
LJ cluster structures that other authors have been point out.

The main concept is to look for a set of cluster's particles
inside of an sphere with  center at the cluster's center of mass with ratio $1.1d^*.$ 

The particles inside of the sphere are natural candidates for being considering the  
cluster's nucleus. 
There many cases, for the selection of the cluster's nucleus. Let $PN$ be the set 
of particles
inside of the sphere, and $cm$ the cluster's coordinates of its center of mass:
\begin{enumerate}
    \item IC or IR when $\exists$ $p_k$ $\in PN$, such that $\arg {k}= min_k ||p_i- cm||,$  
    and $||p_k-cm|| < 0.35 d^*$. By example, $C^*_13, C^*_55, \ldots$ are IC (n1 IC), 
    and $C^*_75, C^*_76, \ldots$ are IR (n1 IR).
    \item IC without a particle as a center when 
    $|PN| =12$ and these $12$ are closed to the sphere's shell. 
    This is the case for IC nucleus with 12 particles (n0 IC). By example, 
    $C^*_521, C^*_533, \ldots$.
    \item When $3 \leq |PN|$ and $|PN \leq 7$, $PN$ is considering the cluster's nucleus. 
    This criteria gives a nucleus with $3$ to $7$ particles. By example, $n3$: 665, 668, 
    672, 673, 728, $\ldots$; $n4$: 26, 86, 87, 88, 89, 90, 91, 92, 93, 94, 95, 98, $\ldots$;
    $n5$: 22, 23, 24, 25, 28, 29, 33, 34, 78, $\ldots$;
 $n6$: 31, 32, 38, 43, 44, 99, 121, $\ldots$;
$n7$: 18, 19, 20, 21, 27, 30, 35, 36, 37, 39, $\ldots$  
   
    \item when $|PN| \geq 8$, adjust the center of the sphere to the center of mass of mass of the 
    the cluster with the $PN$'s particles with 12 neighbors. The new sphere's ratio is set to 
    $0.9 d^*$, then take as the nucleus the particles inside of this new sphere. 
\end{enumerate}

It is showed in the appendix the Matlab routine "S\_plot\_geCl\_LJ.m". 
It  is a version of an algorithm using 
this heuristic. 
Figure~\ref{fig:cl107} depicts the results of the $C^*_{107}$ and its shell with algorithm~\ref{alg:partCluster} with a nucleus defined by the heuristic of this section.

The next algorithm builds a set of coordinates or give a set of number that they correspond to a cluster inside of a region. A region could be and arbitrary set of points, or points of  a lattice, or the points of other big cluster.

\begin{algorithm} ~\label{alg:MatchingCluster} Matching a cluster $C_n$ with a given region $R$

\noindent\textbf{input:} 

$p_n$: array of 3D, for the particles' coordinates of the cluster $C_n$;

$R_k$: array of 3D, with the points' coordinates of the region $R$;

$M$: int, with the number of points of the region $R$;

\noindent \textbf{output:} 

$r[\cdot]$: array of int, for  the corresponding particles' numbers in the $R$;

$s[\cdot]$: array of 3D, for  the closed corresponding particles' coordinates of the points of $R$;

\noindent \textbf{memory}: 

$d_{min}$: real; // minimum distance
$i_{min}$: int; // particle' number
$mk[\cdot]:=0$: array of int, set all to 0;

%\vspace{-1mm}
\noindent\hrulefill
\begin{easylist}
\renewcommand{\labelitemi}{\ }
\setlength\itemsep{-0.05cm}

\item {\bf if} $M < n$ {\bf then}
\item  \msp {\bf print}("Error, it is insufficient the number of points of R for the cluster");
\item \msp {\bf return};
\item {\bf end if}
\item {\bf for} $i:=1$ {\bf to} $n$ {\bf do}
\item  \msp $d_{min} := 10^8$;
\item \msp {\bf for} $k:=1$ {\bf to} $M$ {\bf do}
\item  \msp \msp {\bf if} ($ml[k] == 0$)
{\bf then}
\item   \msp\msp \msp  $d$ $:=$ $dist(p[i], R[k])$;
\item  \msp \msp \msp{\bf if} ($d < d_{min}$) {\bf then}
\item  \msp \msp \msp \msp $d_{min} := d$;
\item  \msp \msp \msp \msp $i_{min} := k$;
\item  \msp \msp \msp{\bf end if}
\item \msp \msp{\bf end if}
\item \msp {\bf end for}
\item \msp $mk[i_{min}] := 1$;
\item \msp $r[i] := i_{min}$;
\item \msp $s[i] := R[i_{min}]$;
\item  {\bf end for}
\item  {\bf return};
\end{easylist}
\end{algorithm}

The previous algorithm always answers with a set of points and with a set on particles' numbers that they correspond to a cluster, but the original and  the output cluster from the region $R$  could have very different shapes. By example, when a cluster is centered at (0,0,0), and the region is box $[20,50]\times [20,50]\times [20,50]$ of the cb lattice. On the other hand, figure~\ref{fig:cl_13_38_inCB} depicts 
$C^*_{13}$ and $C^*_{38}$ inside of a CB region, where the light blue small boxes are their corresponding points of CB lattice.

\begin{figure}
\centerline{ \psfig{figure=\IMAGESPATH/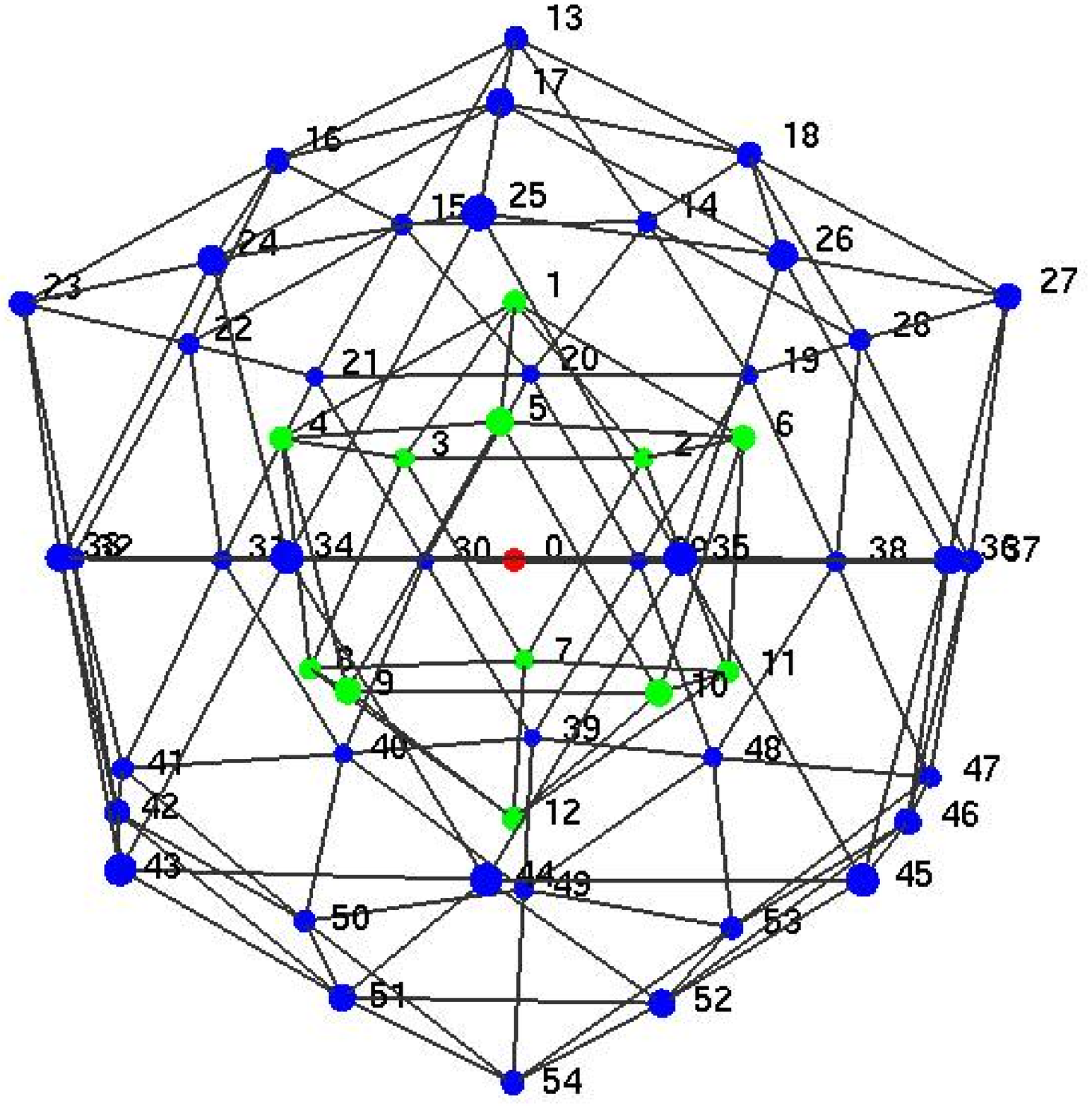,
height=50mm} \psfig{figure=\IMAGESPATH/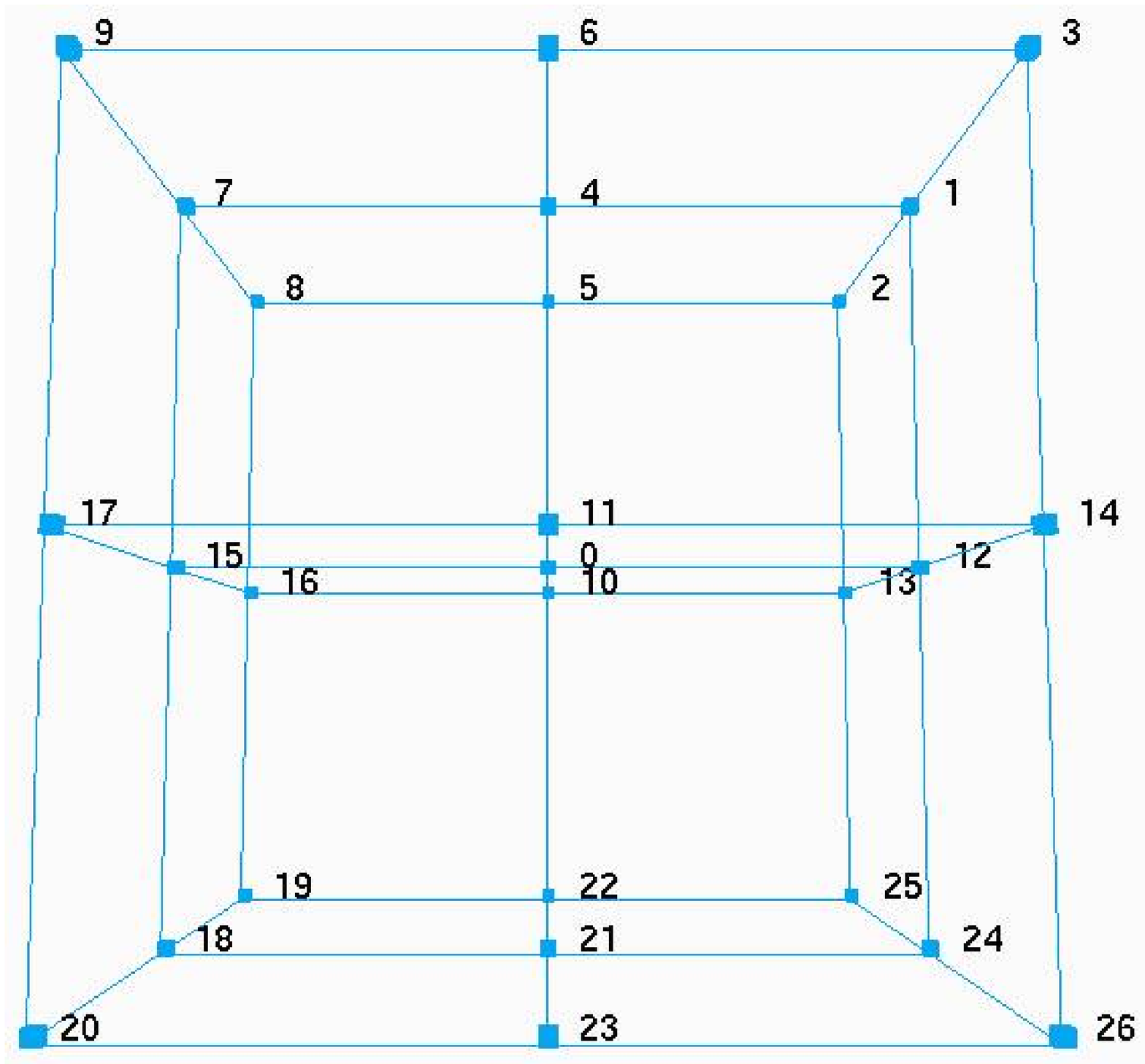,
height=50mm}} \centerline{ \makebox[2.2in][c]{ a)
}\makebox[2.2in][c]{ b) } } \caption{ Examples of numeration from the nucleus to its shell a) region of the IC lattice, and b) region of the CB lattice}~\label{fig:ExamEnu}
\end{figure}

%%%%%%%%%%%%%%%%%%%%%%%%%%%%%%%%%%%%%%%%%%%%%%%%%%%%%%%%%%%%%%%%%%%%%%%%%%%%%%%%%
%%%%%%%%%%%%%%%%%%%%%%%%%%%%%%%%%%%%%%%%%%%%%%%%%%%%%%%%%%%%%%%%%%%%%%%%%%%%%%%%%
\section{Parallel evolutionary algorithm for searching optimal LJ clusters}~\label{sc:parEvAlgorrithm}
%%%%%%%%%%%%%%%%%%%%%%%%%%%%%%%%%%%%%%%%%%%%%%%%%%%%%%%%%%%%%%%%%%%%%%%%%%%%%%%%%
%%%%%%%%%%%%%%%%%%%%%%%%%%%%%%%%%%%%%%%%%%%%%%%%%%%%%%%%%%%%%%%%%%%%%%%%%%%%%%%%%

The novelty of the Parallel evolutionary algorithm of this section is not a new complete paradigm, as I mentioned before, it is a reload version of previous ad-hoc and genetic algorithm (see ~\cite{immas:Gomez1991, aml:Barron1996, aml:Barron1997, aml:Barron1999,
cpc:Romero1999A,
siamAM:Romero1999B,
arXiv:Barron2005, arXiv:Barron2010}).

The term evolutionary algorithm for my approach is justified by the fact that the results are converted into input data, and this cause a change of the expected behavior of the program beyond of its programming. What, I precisely mean is that the efficacy and efficiency of the program for determining optimal LJ clusters is improved. Also, I added new routines based in phenotype and genotype strategies that my previous algorithms have not. 
But in the essence, it is an evolutionary program improved by his creator to increase its efficiency and efficacy with adding changes into its old routines by hand. 
One of the aspect to point out, is that previous version was only based in elitism, here the diversity is favored and it will come from the data of the optimal clusters and for the data of the CB lattice, particularly. 

The algorithm~\cite{cpc:Romero1999A}) now includes the following genotype and phenotype mechanisms.

For the genotype mechanisms, a telephone model for the clusters 
consist to get a set of number to represent a cluster. This can be done with a region of a lattice and a cluster by using the algorithm~\ref{alg:MatchingCluster}.

An enumeration from the nucleus to its shell can be done by ordering the particles of a region by its ratio, y coordinate and its angle on the XY plane. Figure~\ref{fig:ExamEnu} depict a IC and CB regions with this numeration. This helps because for a cluster, lower number are in the core, larger number are in the shell whatever it is respect to a lattice or to other $C^*_n.$

With the telephone model for a cluster, a mutation is like to dial the cluster's telephone with one or more mistakes. The mistake numbers can be replaced by any number not in the cluster's telephone with the condition that such numbers are the indices of particles' coordinates in a given region.

Children can be created by replacing sequences of the clusters' telephones of 2 or more parent clusters.

After the creation of the children by any genetic mechanism, a minimization procedure is applied for the corresponding coordinates of the particles' numbers of the children to get a local minimal LJ cluster for elitism selection.

On the other hand, the previous genetic algorithm adds mechanisms to use the algorithm~\ref{alg:partCluster} with or without the heuristic given in subsection~\ref{scc:heuristic}. New kind of mutations are incorporate by using the matching algorithm~\ref{alg:MatchingCluster} to transform a cluster into a CB, IC, IF or any lattice before crossover and make up.
Previously, the current population include  the optimal LJ cluster, the cluster with more 12 neighbors, and the worst LJ cluster. The change is to include the lower and as many clusters as possible of the 8 categories  of the heuristic for determining the nucleus with the current optimal LJ cluster.

The parrallel algorithm defines a player main routine, which consists in two main routines: Cerberus and Prometheus.

\begin{algorithm} ~\label{alg:Player} Player

\noindent\textbf{input:} 
 
 timer: set an interval of time for sending a stop signal.
 
$I_n$: int parameter of the initial cluster ($\geq 13$);

$F_n$: int parameter of final cluster ($\geq 13$ and ($\geq I_n$));

$P_{sz}$: int parameter of the population size ($\geq 9$).

$R_{CB}$: set of 3D points of the CB lattice ($\gg F_n$);

$R_{L1},\ldots R_{Lk}$: set of 3D points of other lattices ($\gg F_n$);

\noindent \textbf{output:} 

\noindent \textbf{memory}: 

$C^*_{n}$: private data of the current putative optimal LJ clusters;

%\vspace{-1mm}
\noindent\hrulefill
\begin{easylist}
\renewcommand{\labelitemi}{\ }
\setlength\itemsep{-0.05cm}

\item {\bf while} (1) {\bf do}
\item \msp {\bf execute} Cerberus;
\item \msp {\bf execute} Prometheus($I_n,F_n,P_{sz},R_{CB},R_{L1},\ldots R_{Lk}$);
\item \msp {\bf a timer or the user send} a signal to stop;
\item  {\bf end while}
\end{easylist}
\end{algorithm}

Cerberus is an elitism routine for communicating and keeping the best putative optimal LJ clusters. It take care of the communication but never interrupt the process of Prometheus.

\begin{algorithm} ~\label{alg:Cerberus} Cerberus

\noindent\textbf{input:} 

$P_c$: input pile of messages for $C_k$ (LJ clusters);

st$_P$: int, exclusive variable to communicate with Prometheus's state, 1: Prometheus is  searching or 0:Prometheus is not searching;

\noindent \textbf{output:} 

signal: semaphore command for waiting or executing;

\noindent \textbf{memory}: 

$C^*_{n}$: private data of the current putative optimal LJ clusters;

%\vspace{-1mm}
\noindent\hrulefill
\begin{easylist}
\renewcommand{\labelitemi}{\ }
\setlength\itemsep{-0.05cm}

\item {\bf signal} Prometheous {\bf goes};
\item {\bf while} (1) {\bf do}
\item \msp {\bf if} (st$_P$ == 0) {\bf then}
\item \msp \msp {\bf while} pile($P_c$) is not empty {\bf do}
\item \msp \msp {\bf signal} Prometheous {\bf to wait};
 \item \msp \msp $C_k$ := pop($P_c$); 
\item \msp \msp {\bf if} LJ$(C_k)$ $<$ LJ$(C^*_k)$ {\bf then}
\item \msp \msp \msp $(C^*_k)$ = $(C_k)$;
\item \msp \msp \msp {\bf send} message $C^*_k$ to others players;
\item  \msp \msp {\bf end if}
\item  \msp {\bf end while}
\item {\bf signal} Prometheous {\bf to continue};
\item {\bf end while}
\end{easylist}
\end{algorithm}

Prometheus is the implementation of the previous evolutionary algorithm. It has exclusive access to the best LJ clusters during the evolutionary process.

\begin{algorithm} ~\label{alg:Prometheus} Prometheus

\noindent\textbf{input:} 

$I_n$: int parameter of the initial cluster ($\geq 13$);

$F_n$: int parameter of final cluster ($\geq 13$ and ($\geq I_n$));

$P_{sz}$: int parameter of the population size ($\geq 9$).

$R_{CB}$: set of 3D points of the CB lattice ($\gg F_n$);

$R_{L1},\ldots R_{Lk}$: set of 3D points of other lattices ($\gg F_n$);

signal: semaphore command for waiting or executing;

\noindent \textbf{output:} 

st$_P$: int, exclusive variable to communicate its state, 1: busy or 0:waiting;

\noindent \textbf{memory}: 

$C^*_{n}$: private data of the current putative optimal LJ clusters;

%\vspace{-1mm}
\noindent\hrulefill
\begin{easylist}
\renewcommand{\labelitemi}{\ }
\setlength\itemsep{-0.05cm}

\item {\bf while} (1) {\bf do}
\item \msp st$_P$ : = 0;
\item \msp {\bf for} $n:=I_n$ {\bf to} $F_n$ {\bf do}
\item \msp \msp{\bf do} Cerberus(signal) or player(signal);
\item  \msp \msp st$_P$ := 1; 
\item \msp \msp {\bf execute:} evolutionary algorithm for exploring $C_n$
\item \msp \msp st$_P$ :=0; 
\item  \msp {\bf end for}
\item  {\bf end while}
\end{easylist}
\end{algorithm}

When Prometheus is executing the evolutionary algorithm, there is not access to the private memory even for others clusters different of the current $n$. This is because the evolutionary algorithm could use any $C^*_j$ for creating offsprings at any time.  Before or after, the process of the evolutionary algorithm, or when Prometheus is off, there is access to the best LJ optimal clusters.

The player routine is designed for working with copies of it. This could cause a bootle's neck  for the communications. Therefore, it is convenient to define a master player. In this case, only the master has the ability to send and receive messages, meanwhile the slave players can only send messages to it.

\begin{figure}
\centerline{ \psfig{figure=\IMAGESPATH/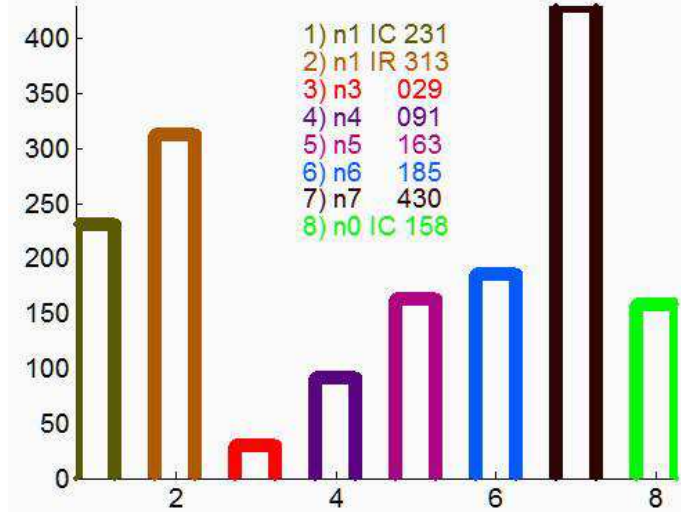,
width=90mm}} \caption{Histogram of nucleus type for $C^*_n,$ 
$n=13,\ldots,1612$ particles}~\label{fig:hist_nucTyp}
\end{figure}

\begin{figure}
\centerline{ \psfig{figure=\IMAGESPATH/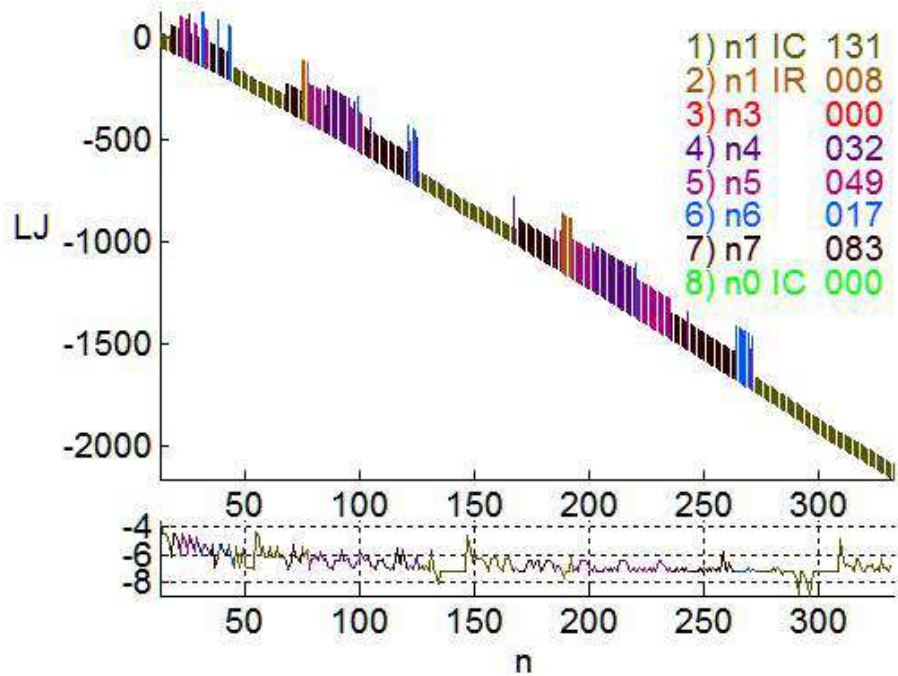,
height=90mm}} \caption{LJ potential, potential difference vs
 $C^*_n,$  $n=13,\ldots,332$ particles}~\label{fig:gtyp13_332}
\end{figure}

\begin{figure}
\centerline{ \psfig{figure=\IMAGESPATH/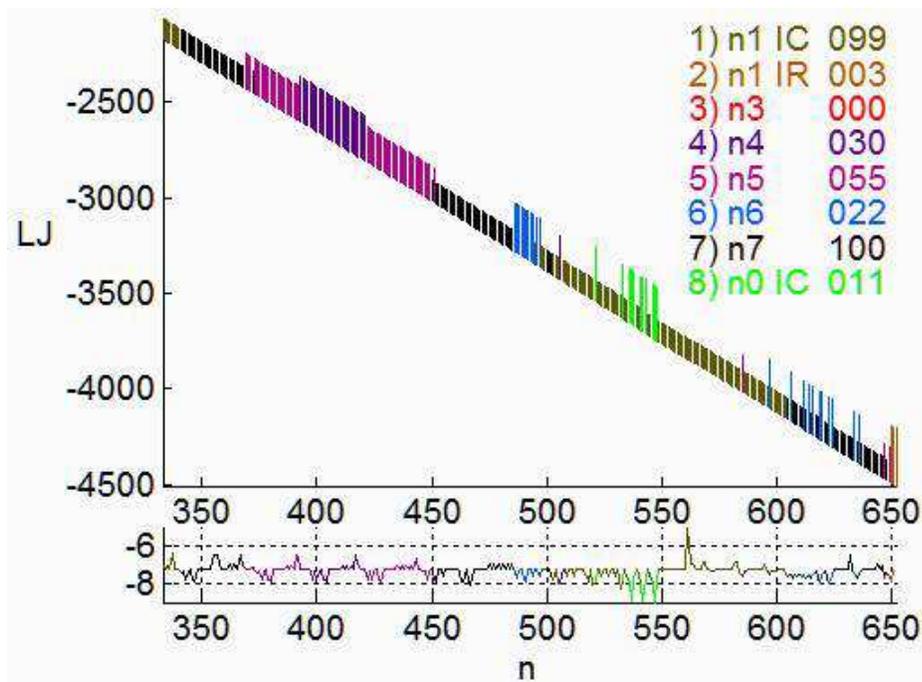,
height=90mm}} \caption{LJ potential, potential difference vs
 $C^*_n,$  $n=333,\ldots,652$ particles}~\label{fig:gtyp333_652}
\end{figure}

\begin{figure}
\centerline{ \psfig{figure=\IMAGESPATH/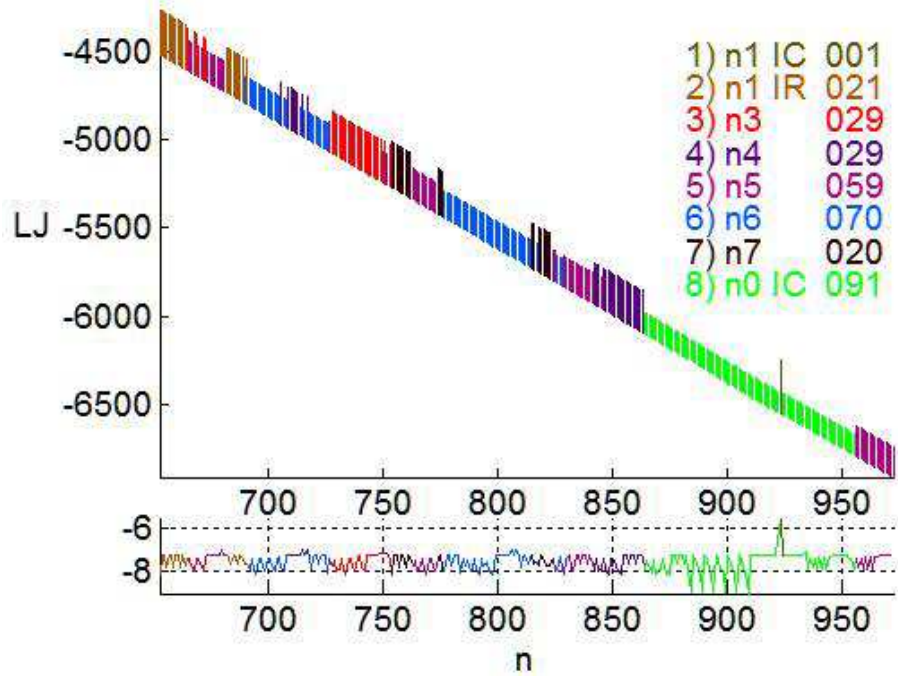,
height=90mm}} \caption{LJ potential, potential difference vs
cluster with $C^*_n,$  $n=653,\ldots,972$ particles}~\label{fig:gtyp653_972}
\end{figure}

\begin{figure}
\centerline{ \psfig{figure=\IMAGESPATH/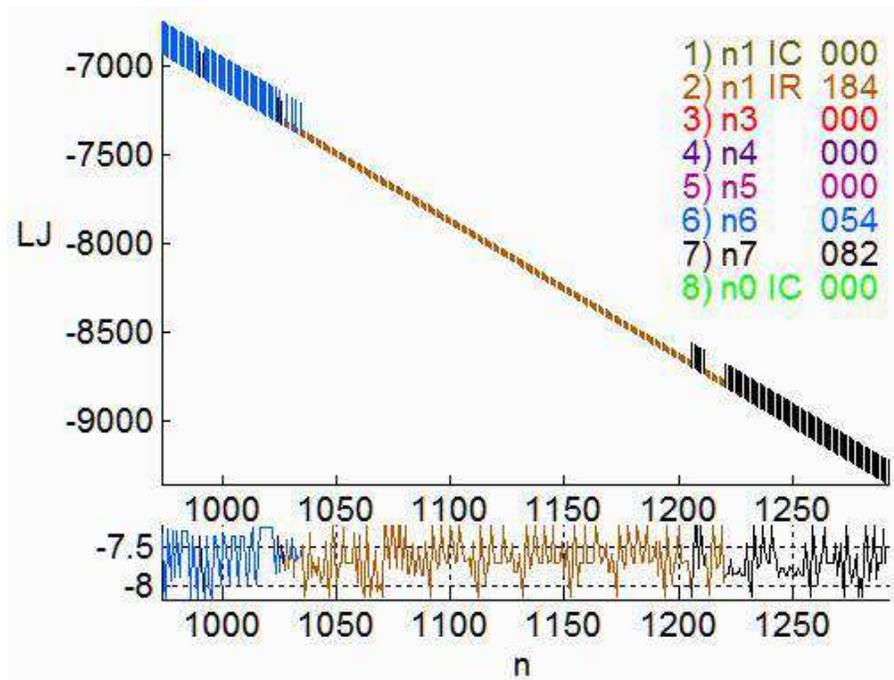,
height=90mm}} \caption{LJ potential, potential difference vs
 $C^*_n,$  $n=973,\ldots,1292$ particles}~\label{fig:gtyp973_1292}
\end{figure}

\begin{figure}
\centerline{ \psfig{figure=\IMAGESPATH/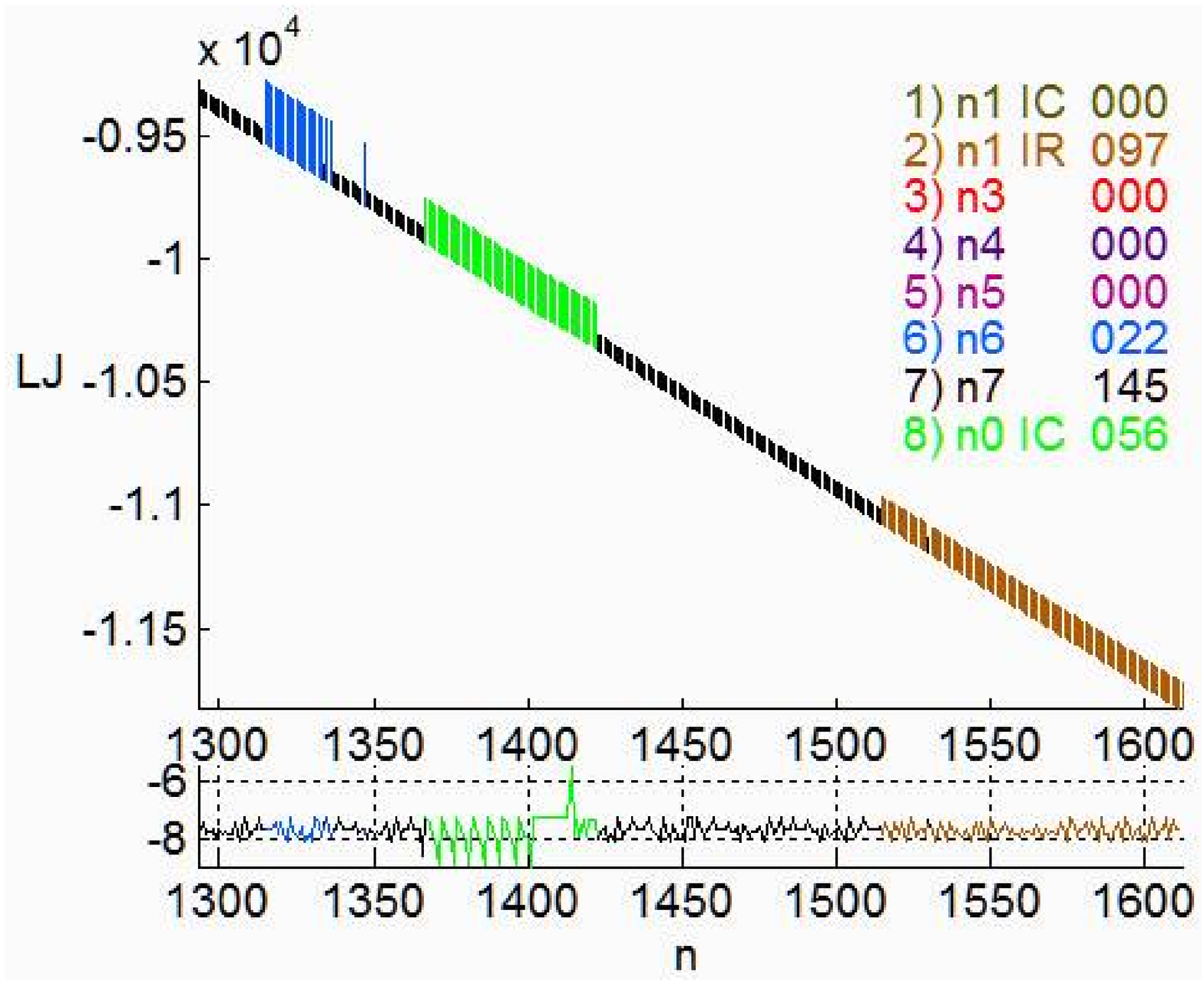,
height=90mm}} \caption{LJ potential, potential difference vs
 $C^*_n,$  $n=1293,\ldots,1612$ particles}~\label{fig:gtyp1293_1612}
\end{figure}

%%%%%%%%%%%%%%%%%%%%%%%%%%%%%%%%%%%%%%%%%%%%%%%%%%%%%%%%%%%%%%%%%%%%%%%%%%%%%%%%%
%%%%%%%%%%%%%%%%%%%%%%%%%%%%%%%%%%%%%%%%%%%%%%%%%%%%%%%%%%%%%%%%%%%%%%%%%%%%%%%%%
%%%%%%%%%%%%%%%%%%%%%%%%%%%%%%%%%%%%%%%%%%%%%%%%%%%%%%%%%%%%%%%%%%%%%%%%%%%%%%%%%
\section{Results}~\label{sc:results}
%%%%%%%%%%%%%%%%%%%%%%%%%%%%%%%%%%%%%%%%%%%%%%%%%%%%%%%%%%%%%%%%%%%%%%%%%%%%%%%%%
%%%%%%%%%%%%%%%%%%%%%%%%%%%%%%%%%%%%%%%%%%%%%%%%%%%%%%%%%%%%%%%%%%%%%%%%%%%%%%%%%
%%%%%%%%%%%%%%%%%%%%%%%%%%%%%%%%%%%%%%%%%%%%%%%%%%%%%%%%%%%%%%%%%%%%%%%%%%%%%%%%%

My previous results~\cite{arXiv:Barron2005} are in the figure~\ref{fig:min_lattice1739}. It depicts a set of particles MIF1739, which 
contains $C^\ast_n$, $n=2,\ldots,1000$. I tried to use MIF1739 has a main lattice from where an algorithm could takes advantage of its building  property:
 $\exists$ $C_n$ $\in$ MIF1739, such that
by  a minimization process,  $C_n$ converges to $C^{\ast}_n$.
However, it is not easy task to locate  a "good"  initial set of points closed to an optimal LJ cluster. Figure~\ref{fig:LJ38:664} depicts where $C^*_{38}$ and $C^*_{664}$ are located into the IF lattice. It is possible to select points by using sphere in MIF1739. Two parameters are need, the ratio and the center of sphere. 

The efficiency of the evolutionary algorithm changes dramatically with the incorporation of the CB lattice, and the phenotype and genotype strategies. The best results comes  from starting with no optimal clusters but by using CB lattice and IC, IR, FC, dodecahedral lattices, and 14 as the size of the population. 

The heuristic for determining a nucleus (see subsection~\ref{scc:heuristic}) and the algorithm~\ref{alg:partCluster}, helped to extend the selection and interaction between clusters to create offspring for mutation or phenotype crossover or genotype crossover. This heuristic classifies into 8 categories by just using the numbers of particles in a nucleus. The nucleus types n4, n5, n6, and n7  are not geometrically equal. A refinement of heuristic is possible, but it has a computation cost.

Figure~\ref{fig:hist_nucTyp} depicts the histogram of 8 categories
resulting of this heuristic for the $C^*_n,$ $n=13, \ldots, 1612.$ Tables~\ref{tb:t1_t2},~\ref{tb:n3_n4},~\ref{tb:n5_n6}, and ~\ref{tb:n7_t8} contains the classification of the clusters. 
It seems that many cases of $C^*_n$ are obtained rapidly by a the make up operation of its previous or next clusters, but also by considering to extend the diversity of the current population to 8 categories. 

My results could help to answer some old conjetures about the morphology of the microclusters. See Hoare~\cite{ap:Hoare1983}: "Werfelmer's essential contribution was to point out the possibility of extremely compact fivefold symmetric structures for N7, suggesting that the pentagonal bipyrimid (N=7) (fig.3(a)) and the icosahedron (N=13) (fig.3(b)) might be the dominant motifs in larger assemblies." It is partially true,  from table~\ref{tb:n7_t8}the type 7), a nucleus with 7 particles (a pentagonal bipyramid) is the dominant motif for $C^*_n$, $n=18,\ldots,1530$, but IC is not a dominant motif from table~\ref{tb:t1_t2}, type 1) and type2) without considering type 8) (a nucleus IC with 12 particles, which was unknown at 1983). 

Figures~\ref{fig:gtyp13_332},~\ref{fig:gtyp333_652}, ~\ref{fig:gtyp653_972},~\ref{fig:gtyp973_1292}, and~\ref{fig:gtyp1293_1612} depict where the different nucleus type appears. At the bottom of each figure the LJ potential difference bet the consecutive clusters is depicted. Some type of optimal clusters are isolated and the LJ potential difference is highly variable from $n$ = 13  to 1420. But after this cluster it seems to diminish its variations for the IR $C^*_n$, $n \geq 142.$ 
My results are 1600 optimal LJ clusters
with $n=13,\ldots, 1612$ particles, (Most of them are posted in The Cambridge Cluster Database (CCD)~\cite{http:CCD}), a novel 8 categories of nucleus classification,
 and   65 new putative LJ Clusters, 
which are not reported at December of 2016. See table~\ref{tb:newCL}.

%%%%%%%%%%%%%%%%%%%%%%%%%%%%%%%%%%%%%%%%%%%%%%%%%%%%%%%%%%%%%%%%%%%%%%%%%%%%%%%%%
%%%%%%%%%%%%%%%%%%%%%%%%%%%%%%%%%%%%%%%%%%%%%%%%%%%%%%%%%%%%%%%%%%%%%%%%%%%%%%%%%
%%%%%%%%%%%%%%%%%%%%%%%%%%%%%%%%%%%%%%%%%%%%%%%%%%%%%%%%%%%%%%%%%%%%%%%%%%%%%%%%%
\section{Conclusions and future work}~\label{sc:conclusions and future work}

The advances of the technology and science of Physics and Chemistry are fantastic, together with the molecular and the nanostructures design. The results presented here have many implications for the computational molecular design and their models and algorithms.

I hope to witness, that it is quite possible to replicate and to improve these results by using one of the top worldwide supercomputer. Some of the definitions are broad, and this research can be easily extended and applied for exploring geometries and interactions of clusters under other molecular potentials.

%%%%%%%%%%%%%%%%%%%%%%%%%%%%%%%%%%%%%%%%%%%%%%%%%%%%%%%%%%%%%%%%%%%%%%%%%%%%%%%%%
%%%%%%%%%%%%%%%%%%%%%%%%%%%%%%%%%%%%%%%%%%%%%%%%%%%%%%%%%%%%%%%%%%%%%%%%%%%%%%%%%
%%%%%%%%%%%%%%%%%%%%%%%%%%%%%%%%%%%%%%%%%%%%%%%%%%%%%%%%%%%%%%%%%%%%%%%%%%%%%%%%%
\section*{Appendix}
%%%%%%%%%%%%%%%%%%%%%%%%%%%%%%%%%%%%%%%%%%%%%%%%%%%%%%%%%%%%%%%%%%%%%%%%%%%%%%%%%
%%%%%%%%%%%%%%%%%%%%%%%%%%%%%%%%%%%%%%%%%%%%%%%%%%%%%%%%%%%%%%%%%%%%%%%%%%%%%%%%%
%%%%%%%%%%%%%%%%%%%%%%%%%%%%%%%%%%%%%%%%%%%%%%%%%%%%%%%%%%%%%%%%%%%%%%%%%%%%%%%%
\appendix{Matlab Programs}

\begin{verbatim}
function  S_plot_Cl_LJ(ncl,xcl,ycl,zcl,nv1,nv2)
% This subroutine draws the geometry of
% a minimal Lennard Jones Potential's cluster
% The input parameters are:
% ncl : numbers of particles
% xcl, ycl, zcl: arrays of numbers corresponding to the 3D
% cluster's coordinates
% nv1, nv2 : integer number to determine the shells to draw
% With nv1=0 and nv2=2, the cluster's nucleus and the first shell
% are depicted (layer 1 is the nucleus).
% ====================================================================
% More information of this subroutine is in the article:
% Discrete Optimal Global Convergence of a Evolutionary Algorithm
% for Clusters under the  Potential of Lennard Jones
% Author: Carlos Barron-Romero
% Universidad Autonoma Metropolitana, campus Azcapotzalco
% Mexico City.
% This subroutine can be freely used, distributed or modified.
% ====================================================================
% Compute center of mass
xclm=mean(xcl);
yclm=mean(ycl);
zclm=mean(zcl);
% Arrays for particle's neighbors and number of neighbors
vec=zeros(ncl,12);
nvc=zeros(ncl,1);
% dmg is the optimal distance of a pair of particles under
% Lennard Jones Potential
dmg=2^(1/6);
% 10% is the factor to define the lower and upper limits
% for accepting a particle's neighbor
tol=0.1;
d_inf = dmg * (1 - tol);
d_sup = dmg * (1 + tol);
%
% This loop determines the particle's neighbor
for i=1:ncl-1
    for j=i+1:ncl
dij=norm([(xcl(i)-xcl(j)),(ycl(i)-ycl(j)),(zcl(i)-zcl(j))]);
        if ((d_inf < dij) && (dij < d_sup))
            nvc(i)=nvc(i)+1;
            vec(i,nvc(i))=j;
            nvc(j)=nvc(j)+1;
            vec(j,nvc(j))=i;
        end
    end
end
% Select a set of particles closed to the cluster's center of mass
ra = dmg*1.1;
pnuc=zeros(13,1);
nnuc=0;
fnuc = 0;
for i=1:ncl
    dcl=norm([(xcl(i)-xclm),(ycl(i)-yclm),(zcl(i)-zclm)]);
    if (dcl < ra)
        nnuc=nnuc+1;
        pnuc(nnuc)=i;
    end
end
% Analyze the set of particles
if (nnuc >0)
    % First case.
    % Look for a particle with 12 neighbors
    % closed to the CM
    inuc = -1;
    d12nuc = 99999.999;
    for k=1:nnuc
        i=pnuc(k);
        if (nvc(i) == 12)
dcl=norm([(xcl(i)-xclm),(ycl(i)-yclm),(zcl(i)-zclm)]);
            if (dcl < d12nuc)
                d12nuc = dcl;
                inuc=i;
            end
        end
    end
    ra_nuc = dmg * 0.35;
    if ((inuc ~= - 1) & (d12nuc < ra_nuc))
        % There is a particle with 12 neighbors
        pnuc(1) = inuc;
        nnuc = 1;
        ra = d12nuc;
        fnuc = 1;
    end
    % After verify that there is no center,
    % Then it is the nucleus with 12 particles
    if ((inuc == - 1) & (nnuc == 12))
        fnuc = 1;
    end
end
% Final case.
% Skip four (tetrahedron, nnuc=4),
% five (Trigonal bipyramid, nnuc=5),
% six particles (octahedron, nnuc=6),
% (pentagonal polyhedron bipyramide, nnuc=7)
%
if ((fnuc == 0) & (nnuc >= 8))
    % Adjust the cluster's center of mass
    % considering only particles with 12 neighbors
    % of the selected set pnuc
    nclm = 1;
    for k=1:nnuc
        i=pnuc(k);
        if (nvc(i) == 12)
            xclm = xclm + xcl(i);
            yclm = yclm + ycl(i);
            zclm = zclm + zcl(i);
            nclm = nclm + 1;
        end
    end
    xclm = xclm / nclm;
    yclm = yclm / nclm;
    zclm = zclm / nclm;
    % Adjust the selected set, keeping the closed
    % to the adjusted center
    ra = dmg*0.9;
    pnuc_nw=zeros(13,1);
    nnuc_nw=0;
    for k=1:nnuc
        i=pnuc(k);
        dcl=norm([(xcl(i)-xclm),(ycl(i)-yclm),(zcl(i)-zclm)]);
        if (dcl < ra)
            nnuc_nw=nnuc_nw+1;
            pnuc_nw(nnuc_nw)=i;
        end
    end
    % Take this new set as the nucleus
    pnuc = pnuc_nw;
    nnuc = nnuc_nw;
    fnuc = 1;
end
% Determine the cluster's layers
% with the nucleus particles
capa=zeros(ncl,1);
ncapa=1;
for jp=1:nnuc
    capa(pnuc(jp))=1;
end
while (1)
    fmk=0;
    for i=1:ncl
        if (capa(i) == ncapa)
            for jv=1:nvc(i)
                pvc=vec(i,jv);
                if (capa(pvc)== 0)
                    capa(pvc) = ncapa+1;
                    fmk=1;
                end
            end
        end
    end
    if (fmk == 0)
        break;
    end
    ncapa = ncapa + 1;
end
% Color table
tclr = [ 1,0,0; ...
    0,1,0; ...
    0,0,1; ...
    1,1,0; ...
    51/255, 153/255, 1; ...
    1,0,1; ...
    0.5,0.5,1; ...
    0.5,0.5,0.5; ...
    1,0.5,1; ...
    1,0.5,0.5; ...
    1,0.5,0.75; ...
    1,0.75,0; ...
    0,0,0];
hold on;
for i=1:ncl
    if (capa(i) < nv1)
        continue;
    end
    if (capa(i) > nv2)
        continue;
    end
    % cluster's lines
    for jv=1:nvc(i)
        pvc=vec(i,jv);
        if (capa(i) == capa (pvc))
            line([xcl(i),xcl(pvc)], ...
                [ycl(i),ycl(pvc)], ...
                [zcl(i),zcl(pvc)], ...
                'Color',tclr(capa(i),:), ...
                'LineWidth',1);

        end
    end
    % particles
    plot3(xcl(i), ...
        ycl(i), ...
        zcl(i), '-ko',...
        'LineWidth',1,...
        'MarkerEdgeColor','k', ...
        'MarkerFaceColor',tclr(capa(i),:),...
        'MarkerSize',8);
end
axis equal;
grid on;
linp=sprintf('Cluster of %d particles',ncl);
title(linp);
view(45,45);
hold off;
\end{verbatim}

\begin{verbatim}
function [n,x,y,z] = S_read_Cl_LJ(file_name)
% This subroutine reads in the format of the file from the
% The Cambridge Energy Landscape Database (http://www-wales.ch.cam.ac.uk/CCD.html)
% of a minimal cluster under Lennard Jones Potential
%
% The input parameters is the file name
% Is is ####.txt, where #### is the cluster's number of particles
%
% ====================================================================
% This subroutine is in the article:
% Discrete Optimal Global Convergence of a Evolutionary Algorithm
% for Clusters under the  Potential of Lennard Jones
% Author: Carlos Barron-Romero
% Universidad Autonoma Metropolitana, campus Azcapotzalco
% Mexico City.
% This subroutine can be freely used, distributed or modified under
% your own responsibility.
% ====================================================================
%
fid=fopen(file_name);
p = fscanf(fid,'%g %g %g', [3 inf]);
fclose(fid);
% x, y and z are the cluster's coordinates
x=p(1,:);
y=p(2,:);
z=p(3,:);
% n is the cluster's number of particles
n=length(x);
\end{verbatim}

\begin{verbatim}
% This program calls the
% subroutine S_plot_geCl_LJ_cl to draw the geometry of
% an optimal Lennard Jones cluster
% ====================================================================
% More information of this subroutine is in the article:
% Discrete Optimal Global Convergence of a Evolutionary Algorithm
% for Clusters under the  Potential of Lennard Jones
% Author: Carlos Barron-Romero
% Universidad Autonoma Metropolitana, campus Azcapotzalco
% Mexico City.
% This subroutine can be freely used, distributed or modified under
% your own responsibility.
% ====================================================================
feature('UseGenericOpenGL',0);
[filename,pname] = uigetfile('*.TXT');
% Define your own routine to read the particles' coordinates
% of a optimal cluster under the Leenard Jones Potential
% or S_read_cl_LJ is set to read the files in the
% The Cambridge Energy Landscape Database (http://www-wales.ch.cam.ac.uk/CCD.html)
%
% The input parameters is the file name
% Is is ####.txt, where #### is the cluster's number of particles
[ncl, xcl,ycl,zcl] = S_read_Cl_LJ([pname,filename]);
nv1=0;
nv2=2;
clf;
S_plot_geCl_LJ(ncl,xcl,ycl,zcl,nv1,nv2);

\end{verbatim}

\begin{table}
\centerline{\small
\begin{tabular*}{5in}[t]{p{1.6in} | p{3.1in}}
{\bf $\mathbf{n1}$ IC } \ \ 231 &   {\bf $\mathbf{n1}$ IR}  \ \ 313  \\ 
\hline \hline
%%%%%%%%%%%%%%%%%%%%%%%%%%%%%%%%%%%%%%%%%%%%%%%%%%%%%%%%%%%%
13  14  15  16  17  45  46  47  48  49  50  51  52  53  54  55  56 57  58  59  60  61  62  63  64  65  66  67  126  127  128  129 130 131  132  133  134  135  136  137  138  139  140  141  142 143 144 145  146  147  148  149  150  151  152  153  154  155 156 157  158 159  160  161  162  163  164  165  166  168  272 273 274  275  276 277  278  279  280  281  282  283  284  285 286 287  288  289  290 291  292  293  294  295  296  297  298 299 300  301  302  303  304 305  306  307  308  309  310  311 312 313  314  315  316  317  318 319  320  321  322  323  324 325 326  327  328  329  330  331  332 333  334  335  336  337 338 339  340  495  498  499  503  504  505 507  508  509  510 511 512  513  514  515  516  517  518  519  520 522  523  524 525 526  527  528  529  530  531  532  534  535  539 540  544 545 549  550  551  552  553  554  555  556  557  558  559 560 561 562  563  564  565  566  567  568  569  570  571  572  573 574 575  576  577  578  579  580  581  582  583  584  586  587 588 589  590  591  592  593  594  595  596  598  599  600  601 602 603 604  923 & 
75  76  77  188  189  190  191  192  650  651  652  653  654  655 656  657  658  659  660  661  662  663  664  682  683  684  685 686  687  688  689  691  1027  1029  1031  1033  1035  1036  1037 1038  1039  1040  1041  1042  1043  1044  1045  1046  1047  1048 1049  1050  1051  1052  1053  1054  1055  1056  1057  1058  1059 1060  1061  1062  1063  1064  1065  1066  1067  1068  1069  1070 1071  1072  1073  1074  1075  1076  1077  1078  1079  1080  1081 1082  1083  1084  1085  1086  1087  1088  1089  1090  1091  1092 1093  1094  1095  1096  1097  1098  1099  1100  1101  1102  1103 1104  1105  1106  1107  1108  1109  1110  1111  1112  1113  1114 1115  1116  1117  1118  1119  1120  1121  1122  1123  1124  1125 1126  1127  1128  1129  1130  1131  1132  1133  1134  1135  1136 1137  1138  1139  1140  1141  1142  1143  1144  1145  1146  1147 1148  1149  1150  1151  1152  1153  1154  1155  1156  1157  1158 1159  1160  1161  1162  1163  1164  1165  1166  1167  1168  1169 1170  1171  1172  1173  1174  1175  1176  1177  1178  1179  1180 1181  1182  1183  1184  1185  1186  1187  1188  1189  1190  1191 1192  1193  1194  1195  1196  1197  1198  1199  1200  1201  1202 1203  1204  1205  1212  1213  1214  1215  1216  1217  1218  1219 1220  1515  1516  1517  1518  1519  1520  1521  1522  1523  1524 1525  1526  1527  1528  1529  1531  1532  1533  1534  1535  1536 1537  1538  1539  1540  1541  1542  1543  1544  1545  1546  1547 1548  1549  1550  1551  1552  1553  1554  1555  1556  1557  1558 1559  1560  1561  1562  1563  1564  1565  1566  1567  1568  1569 1570  1571  1572  1573  1574  1575  1576  1577  1578  1579  1580 1581  1582  1583  1584  1585  1586  1587  1588  1589  1590  1591 1592  1593  1594  1595  1596  1597  1598  1599  1600  1601  1602 1603  1604  1605  1606  1607  1608  1609  1610  1611  1612 \\ \hline
%%%%%%%%%%%%%%%%%%%%%%%%%%%%%%%%%%%%%%%%%%%%%%%%%%%%%%%%%%%%
\end{tabular*}
}
\caption{Type 1 and 2 of the $C^*_n, n=13,\ldots,1612$ }~\label{tb:t1_t2}
\end{table}

\begin{table}
\centerline{\small
\begin{tabular*}{5in}[t]{p{1.1in} | p{3.6in}}
{\bf $\mathbf{n3}$}  \ \ 29 &   {\bf $\mathbf{n4}$}   \ \ 91 \\
\hline \hline
%%%%%%%%%%%%%%%%%%%%%%%%%%%%%%%%%%%%%%%%%%%%%%%%%%%%%%%%%%%%
665  668  669  672  673  728  729  730  731  732  733  734  735 736  737  738  739  740  741  742  743  744  745  746  747  748 749  751  
753 &
26 86 87 88 89 90 91 92 93 94 95 98 125 167 201 203 204 205 206 207 208 209 210 211 212 213 214 215 216 217 218 219 393 394 395 396 397 398 399 400 401 402 403 404 405 406 407 408 409 410 411 412 413 414 415 416 417 418 419 420 421 506 706 709 710 711 712 713 715 717 842 843 844 846 847 848 849 850 851 852 853 854 855 856 857 858 859 860 861 862 863
\\ \hline
%%%%%%%%%%%%%%%%%%%%%%%%%%%%%%%%%%%%%%%%%%%%%%%%%%%%%%%%%%%%
\end{tabular*}
}
\caption{Type 3 and 4 of the $C^*_n, n=13,\ldots,1612$}~\label{tb:n3_n4}
\end{table}

\begin{table}
\centerline{\small
\begin{tabular*}{5in}[t]{p{2.0in} | p{2.7in}}
{\bf $\mathbf{n5}$}  \ \ 163 &   {\bf $\mathbf{n6}$} \ \ 185 \\
\hline \hline
%%%%%%%%%%%%%%%%%%%%%%%%%%%%%%%%%%%%%%%%%%%%%%%%%%%%%%%%%%%%
22 23 24 25 28 29 33 34 78 79 80 81 82 83 84 96 97 100 101 105 122 185 187 193 194 195 196 197 198 199 200 202 221 222 223 224 225 226 227 228 229 230 231 232 233 234 235 243 270 369 370 371 373 374 375 376 377 378 379 380 381 382 383 384 385 386 387 388 389 390 391 392 422 423 424 425 426 427 428 429 430 431 432 433 434 435 436 437 438 439 440 441 442 443 444 445 446 447 448 449 451 585 647 649 666 667 670 671 674 675 676 677 678 679 680 681 727 750 752 763 764 765 766 767 768 769 770 771 772 773 824 825 829 830 831 832 833 834 835 836 837 838 839 840 841 845 956 957 958 959 960 961 962 963 964 965 966 967 968 969 970 971 972 &
31 32 38 43 44 99 121 123 124 220 264 265 266 267 268 269 271 486 487 488 489 490 491 492 493 494 496 497 597 606 612 614 616 619 620 623 624 634 636 690 692 693 694 695 696 697 698 699 700 701 702 703 704 705 707 708 714 716 718 719 720 721 722 723 724 725 726 777 778 779 780 781 782 783 784 785 786 787 788 789 790 791 792 793 794 795 796 797 798 799 800 801 802 803 804 805 806 807 808 809 810 811 812 813 814 817 826 827 828 973 974 975 976 977 978 979 980 981 982 983 984 985 986 987 988 989 992 993 994 995 996 997 998 999 1000 1001 1002 1003 1004 1005 1006 1007 1008 1009 1010 1011 1012 1013 1014 1015 1016 1017 1018 1019 1020 1021 1022 1023 1025 1028 1030 1032 1034 1315 1316 1317 1318 1319 1320 1321 1322 1323 1324 1325 1326 1327 1328 1329 1330 1331 1332 1333 1335 1336 1347
\\ \hline
%%%%%%%%%%%%%%%%%%%%%%%%%%%%%%%%%%%%%%%%%%%%%%%%%%%%%%%%%%%%
\end{tabular*}
}
\caption{Type 5 and 6 of the $C^*_n, n=13,\ldots,1612$}~\label{tb:n5_n6}
\end{table}

\begin{table}
\centerline{\small
\begin{tabular*}{5in}[t]{p{3.4in} | p{1.3in}}
{\bf $\mathbf{n7}$} \ \ 430 &   {\bf $\mathbf{n0}$ IC} \ \ 158\\
\hline \hline
%%%%%%%%%%%%%%%%%%%%%%%%%%%%%%%%%%%%%%%%%%%%%%%%%%%%%%%%%%%%
18 19 20 21 27 30 35 36 37 39 40 41 42 68 69 70 71 72 73 74 85 102 103 104 106 107 108 109 110 111 112 113 114 115 116 117 118 119 120 169 170 171 172 173 174 175 176 177 178 179 180 181 182 183 184 186 236 237 238 239 240 241 242 244 245 246 247 248 249 250 251 252 253 254 255 256 257 258 259 260 261 262 263 341 342 343 344 345 346 347 348 349 350 351 352 353 354 355 356 357 358 359 360 361 362 363 364 365 366 367 368 372 450 452 453 454 455 456 457 458 459 460 461 462 463 464 465 466 467 468 469 470 471 472 473 474 475 476 477 478 479 480 481 482 483 484 485 500 501 502 605 607 608 609 610 611 613 615 617 618 621 622 625 626 627 628 629 630 631 632 633 635 637 638 639 640 641 642 643 644 645 646 648 754 755 756 757 758 759 760 761 762 774 775 776 815 816 818 819 820 821 822 823 990 991 1024 1026 1206 1207 1208 1209 1210 1211 1221 1222 1223 1224 1225 1226 1227 1228 1229 1230 1231 1232 1233 1234 1235 1236 1237 1238 1239 1240 1241 1242 1243 1244 1245 1246 1247 1248 1249 1250 1251 1252 1253 1254 1255 1256 1257 1258 1259 1260 1261 1262 1263 1264 1265 1266 1267 1268 1269 1270 1271 1272 1273 1274 1275 1276 1277 1278 1279 1280 1281 1282 1283 1284 1285 1286 1287 1288 1289 1290 1291 1292 1293 1294 1295 1296 1297 1298 1299 1300 1301 1302 1303 1304 1305 1306 1307 1308 1309 1310 1311 1312 1313 1314 1334 1337 1338 1339 1340 1341 1342 1343 1344 1345 1346 1348 1349 1350 1351 1352 1353 1354 1355 1356 1357 1358 1359 1360 1361 1362 1363 1364 1365 1366 1423 1424 1425 1426 1427 1428 1429 1430 1431 1432 1433 1434 1435 1436 1437 1438 1439 1440 1441 1442 1443 1444 1445 1446 1447 1448 1449 1450 1451 1452 1453 1454 1455 1456 1457 1458 1459 1460 1461 1462 1463 1464 1465 1466 1467 1468 1469 1470 1471 1472 1473 1474 1475 1476 1477 1478 1479 1480 1481 1482 1483 1484 1485 1486 1487 1488 1489 1490 1491 1492 1493 1494 1495 1496 1497 1498 1499 1500 1501 1502 1503 1504 1505 1506 1507 1508 1509 1510 1511 1512 1513 1514 1530 &
521 533 536 537 538 541 542 543 546 547 548 864 865 866 867 868 869 870 871 872 873 874 875 876 877 878 879 880 881 882 883 884 885 886 887 888 889 890 891 892 893 894 895 896 897 898 899 900 901 902 903 904 905 906 907 908 909 910 911 912 913 914 915 916 917 918 919 920 921 922 924 925 926 927 928 929 930 931 932 933 934 935 936 937 938 939 940 941 942 943 944 945 946 947 948 949 950 951 952 953 954 955 1367 1368 1369 1370 1371 1372 1373 1374 1375 1376 1377 1378 1379 1380 1381 1382 1383 1384 1385 1386 1387 1388 1389 1390 1391 1392 1393 1394 1395 1396 1397 1398 1399 1400 1401 1402 1403 1404 1405 1406 1407 1408 1409 1410 1411 1412 1413 1414 1415 1416 1417 1418 1419 1420 1421 1422
\\ \hline
%%%%%%%%%%%%%%%%%%%%%%%%%%%%%%%%%%%%%%%%%%%%%%%%%%%%%%%%%%%%
\end{tabular*}
}
\caption{Type 7 and 8 of the $C^*_n, n=13,\ldots,1612$}~\label{tb:n7_t8}
\end{table}

\begin{table}
\centerline{\small
\begin{tabular*}{5in}[t]{p{4.9in}}
{\bf ($n$, LJ$_{old}$, LJ$_{new}$)}  \ \ 65\\
\hline \hline
%%%%%%%%%%%%%%%%%%%%%%%%%%%%%%%%%%%%%%%%%%%%%%%%%%%%%%%%%%%%
(293,-1888.4271,-1888.4274)
(506,-3427.6212,-3427.6875)
(521,-3539.3314,-3539.5098)
(533,-3628.2529,-3629.2999)
(662,-4581.2049,-4581.2058)
(664,-4596.1971,-4596.1978)
(813,-5712.2507,-5712.2517)
(974,-6928.5630,-6928.6305)
(1064,-7616.1673,-7616.1680)
(1075,-7700.8750,-7700.8755)
(1102,-7905.8577,-7905.8651)
(1103,-7913.5579,-7913.5631)
(1106,-7935.9638,-7935.9689)
(1115,-8004.9485,-8004.9868)
(1125,-8081.1711,-8081.1859)
(1126,-8088.8631,-8088.8764)
(1143,-8218.2614,-8218.2690)
(1144,-8225.9422,-8225.9630)
(1146,-8240.8643,-8240.8662)
(1147,-8248.5659,-8248.5706)
(1148,-8256.2554,-8256.2609)
(1158,-8333.0767,-8333.0809)
(1161,-8355.7255,-8355.7298)
(1162,-8363.4124,-8363.4313)
(1163,-8371.1178,-8371.1218)
(1166,-8393.5692,-8393.5838)
(1167,-8401.2616,-8401.2743)
(1179,-8493.0215,-8493.0221)
(1184,-8530.6521,-8530.6600)
(1185,-8538.3320,-8538.3532)
(1187,-8553.2573,-8553.2634)
(1189,-8568.6507,-8568.6566)
(1225,-8844.5625,-8844.5758)
(1243,-8982.2245,-8982.2304)
(1244,-8989.9194,-8989.9244)
(1275,-9229.3690,-9229.3694)
(1287,-9322.0063,-9322.0069)
(1289,-9337.6149,-9337.6157)
(1292,-9360.2960,-9360.2969)
(1294,-9375.7081,-9375.7091)
(1312,-9515.0285,-9515.0327)
(1315,-9537.6838,-9537.6880)
(1317,-9553.0747,-9553.0788)
(1324,-9606.7794,-9606.7798)
(1336,-9699.4110,-9699.4116)
(1338,-9715.0285,-9715.0314)
(1341,-9737.7061,-9737.7094)
(1343,-9753.1169,-9753.1200)
(1366,-9930.4757,-9930.4764)
(1445,-10538.9878,-10538.9883) \\
(1457,-10631.4452,-10631.4458) 
(1483,-10831.4787,-10831.4791) \\
(1488,-10869.6845,-10869.6917)
(1489,-10877.3862,-10877.4015) \\
(1490,-10885.0944,-10885.1024)
(1523,-11139.7180,-11139.7228) \\
(1526,-11162.5614,-11162.5647)
(1528,-11178.2283,-11178.2291)\\
(1552,-11364.4027,-11364.4088)
(1554,-11380.0717,-11380.0735)\\
(1558,-11410.7180,-11410.7193)
(1585,-11620.2304,-11620.2305)\\
(1595,-11698.0815,-11698.0820)
(1611,0.0000,-11821.7581)
(1612,0.0000,-11829.4494)
\\ \hline
%%%%%%%%%%%%%%%%%%%%%%%%%%%%%%%%%%%%%%%%%%%%%%%%%%%%%%%%%%%%
\end{tabular*}
}
\caption{New $C^*_n$}~\label{tb:newCL}
\end{table}

%\bibliographystyle{abbrv}
%\bibliography{\BIBPATH/NPComplexity_v16}

\end{document}